\documentclass[pdftex,twocolumn,epjc3]{svjour3}
\pdfoutput=1
\RequirePackage[T1]{fontenc}
\RequirePackage{graphicx}
\RequirePackage{mathptmx}      
\RequirePackage{flushend}
\RequirePackage[colorlinks,citecolor=blue,urlcolor=blue,linkcolor=blue]{hyperref}
\journalname{Eur. Phys. J. C}

\begin{document}

\title{Reconstruction of physics objects at the Circular Electron Positron Collider with Arbor}
\author{
      Manqi Ruan\thanksref{e1, addr1}
      \and
      Hang Zhao\thanksref{addr1}
      \and
      Gang Li\thanksref{addr1}
      \and
      Chengdong Fu\thanksref{addr1}
      \and
      Zhigang Wang\thanksref{addr1}
      \and
      Xinchou Lou\thanksref{addr1}
      \and
      Dan Yu\thanksref{addr1, addr2} 
      \and
      Vincent Boudry\thanksref{addr2}
      \and
      Henri Videau\thanksref{addr2}
      \and
      Vladislav Balagura\thanksref{addr2}
      \and
      Jean-Claude Brient\thanksref{addr2}
      \and
      Peizhu Lai\thanksref{addr3}
      \and
      Chia-Ming Kuo\thanksref{addr3}
      \and
      Bo Liu\thanksref{addr1, addr4}
      \and
      Fenfen An\thanksref{addr1, addr4}
      \and
      Chunhui Chen\thanksref{addr4}
      \and
      Soeren Prell\thanksref{addr4}
      \and
      Bo Li\thanksref{addr5}
      \and
      Imad Laketineh\thanksref{addr5}
}

\thankstext{e1}{e-mail: Manqi.ruan@ihep.ac.cn}
\institute{Institute of High Energy Physics, Beijing\label{addr1}
          \and
          Laboratoire Leprince-Ringuet, Ecole Polytechnique, Palaiseau\label{addr2}     
          \and
          National Central University, Taoyuan City\label{addr3}
          \and
          Iowa State University, Ames\label{addr4}
          \and
          Institute de Physique Nucleaire de Lyon, Lyon\label{addr5}
}

\maketitle

\begin{abstract}

After the Higgs discovery, precise measurements of the Higgs properties and the electroweak observables become vital for the experimental particle physics. 
A powerful Higgs/Z factory, the Circular Electron Positron Collider (CEPC) is proposed.
The Particle Flow oriented detector design is proposed to the CEPC and a Particle Flow algorithm, Arbor is optimized accordingly. 
We summarize the physics object reconstruction performance of the Particle Flow oriented detector design with Arbor algorithm and conclude that this combination fulfills the physics requirement of CEPC. 

\end{abstract}

\section{Introduction}
\label{intro}

\subsection {The Higgs discovery and the precision measurements }
\label{Motivation}

The discovery of the Higgs boson completes the entire Standard Model (SM) particle spectrum~\cite{ATLASHiggsDiscovery}\cite{CMSHiggsDiscovery}.
As one of the most successful models that mankind ever constructed, the SM agrees with, predicts and interprets almost all the data taken from the collider experiments.
However, the SM is incapable to explain lots of observed or anticipated fundamental phenomena beyond the collider experiments. 
For instance, the SM consists of no candidate particle for the dark matter, it cannot explain the dark energy and inflation, and so far it doesn't provide enough CP violation for the baryogenesis. 
In addition, the SM suffers from the problem of the naturalness, the hierarchy, and the vacuum stability, {\it etc}.
All these clues point to an intriguing, and highly probable possibility: the SM is a low-energy effective theory of much profound physics principles.
The revelation of these principles is the key objective of experimental particle physics after the Higgs discovery, or say, in the Post-Higgs era. 

Interestingly, most of the clues point to the Higgs field.
The huge difference between the Higgs boson mass and the Planck scale stands for the naturalness problem; 
the couplings between Higgs boson and the SM fermions inhabit the CP violation phases.
The Higgs boson may serve as a portal to the dark matter and even dark energy.
Therefore, the Higgs boson is an excellent probe towards these fundamental physics principles, 
and a Higgs factory that can reveal the nature of the Higgs boson become a must for the experimental particle physics. 

The LHC is a powerful Higgs factory.
It not only discovers the Higgs boson but also indicates the discovered Higgs boson is highly SM-like~\cite{PDG-Higgs}. 
The planned high-luminosity operation of the LHC (HL-LHC) will certainly shed more light on the nature of the Higgs boson. 
However, at a proton collider, the accuracies of the Higgs measurements are limited by the huge QCD background, and most of the Higgs signals can only be identified from its decay final state.
As a result, a very small fraction (roughly $10^{-3}$) of the Higgs events are identified at the proton collider.
The measurement precision (i.e. the signal strengths) is typically limited to 10\% level at the HL-LHC~\cite{ATLAS-HL-LHC}\cite{CMS-HL-LHC}.

The electron-positron collider provides crucial information on top of the HL-LHC. 
First of all, the electron-positron Higgs factory is free of the QCD background.
Within the detector fiducially volume, the ratio between the Higgs signal cross section and that of the inclusive physics events is roughly $10^{-2} \sim 10^{-3}$,
roughly eight orders of magnitude better than the LHC.
The entire event rate at an electron-positron Higgs factory is so low that almost every physics event could be recorded. 
In addition, a significant portion of the Higgs boson is generated with a $Z$ boson (the Higgsstrahlung process) at an electron-positron Higgs factory.
At these events, the Higgs boson could be identified through the $Z$ boson via the recoil mass method,
leading to absolute measurements of the inclusive $ZH$ cross section, Higgs boson width and couplings between the Higgs boson to its decay final states.
The electron-positron collider is also extremely sensitive to the exotic Higgs decay mode search.

For these advantages, many electron-positron Higgs factories have been proposed~\cite{ILC}\cite{CLIC}\cite{FCCee}\cite{CEPCPreCDR}.
The fact that the Higgs boson has 125 GeV mass promotes the concepts of circular Higgs factories, which is upgradable to high energy proton colliders.
The Circular Electron-Positron Collider (CEPC) is one of these proposals. 
With a main ring circumference of 100 km, the CEPC will be operated at 240~GeV center of mass energy and produce 1 million Higgs boson in 10 years' operation with two detectors.
At this energy, roughly 95\% of the Higgs bosons are generated via the $ZH$ process, ensuring an excellent $g(HZZ)$ measurement.
Lowing the center of mass energy to 91~GeV, the CEPC could produce more than $10^{10}$ $Z$ boson per year.
From which, electroweak observables such as $A_FB^b$, $R_b$, the Z line shape can be measured precisely.
After the electron-positron collision phase, a super proton proton collider (SppC) with a center of mass energy up to 100 TeV can be installed in the same tunnel.

In terms of the Higgs measurement, the CEPC determines the absolute Higgs couplings to accuracies of 0.1\% - 1\%, roughly one order of magnitude superior to the model dependent measurements at the HL-LHC ~\cite{ATLAS-HL-LHC}\cite{CMS-HL-LHC}. 
The Higgs total width could be measured to an accuracy of 3\%.
Depends on the event topology, the exotic decay branching ratios can be limited to $10^{-3}$ to $10^{-5}$~\cite{HaoZhangExotic}.
Meanwhile, the CEPC produces lots of $Z$ and $W$ bosons, it can boost the precisions of EW measurements by at least one order of magnitude from current precision. 
A combination of the electroweak (EW) and the Higgs measurements could significantly enhance the physics reach~\cite{JiaYinCombination}.

\subsection { The CEPC physics requirements and the Particle Flow}
\label{DetectorReq}

As a Higgs factory, the CEPC detector should be able to distinguish the Higgs signal from the SM background and to classify different Higgs generation/decay modes.
In another word, the CEPC detector is required to reconstruct all the physics objects in the Higgs events with high efficiency, high purity and measure them with high precision.
The physics requirements for the CEPC detector could be schematized (but not limited to) as follows:

\begin{itemize}

\item []1, Be adequate to the CEPC collision environment: the detector should be fast enough to record all the physics events and robust enough against the irradiation. 

\vspace*{0.3cm}

\item []2, Highly hermetic;

\vspace*{0.3cm}

\item []3, Excellent track reconstruction efficiency and momentum resolution better than $\delta(\frac{1}{P_t}) = 2\times 10^{-5}(\mbox{GeV}^{-1})$, required by {\it $g(H\mu^+\mu^-)$} 
measurement and the Higgs recoil mass reconstruction at $llH$ channels;

\vspace*{0.3cm}

\item []4, Excellent lepton identification, required by both Higgs measurements and EW measurements;

\vspace*{0.3cm}

\item []5, Capable to identify charged kaons, required by the flavor physics;

\vspace*{0.3cm}

\item []6, Precise reconstruction of photons, required by physics with $\tau$ final states, jet energy reconstruction, and the $Br(H\to \gamma\gamma)$ measurement;

\vspace*{0.3cm}

\item []7, Capable to identify $\tau$ lepton and different decay modes of the $\tau$ lepton, requested by $g(H\tau^+\tau^-)$ measurements and physics with $\tau$ final states;

\vspace*{0.3cm}

\item []8, Good Jet/Missing Energy (MET) reconstruction, appreciated by most of the CEPC physics measurements;

\vspace*{0.3cm}

\item []9, Capable to separate $b$-jets, $c$-jets and light jets ($uds$ and gluon jets): required by the $g(Hb\bar{b})$, $g(Hc\bar{c})$, and $g(Hgg)$ measurements.

\end{itemize}

Since the $W$ and $Z$ bosons decay into similar physics objects as the Higgs boson, the EW measurements also benefit from these requirements.
In addition, compared to the Higgs measurements, the EW measurements are much demanding in the systematic control.
For example, the CEPC detector is required to determine the luminosity to a relative accuracy of $10^{-3}$ for the Higgs measurements, and $10^{-4}$ for the $Z$ pole operation.

Adequate reconstruction and detector design are fundamental to the CEPC. 
As a significant trend for the experimental particle physics~\cite{eepfa}\cite{CALICEReview}\cite{CMS-HGC}\cite{ATLAS-HPTD}, the Particle Flow oriented detector design and reconstruction is selected as the baseline for the CEPC. 
The Particle Flow aims at reconstructing all the final state particles with the most suited sub-detector system.
Ultimately, it provides 1-1 correspondence between the reconstructed particles and the physics truth. 
The physics objects are then reconstructed from the final state particles.
The Particle Flow, with an adequate detector design, can significantly enhance the reconstruction efficiency, purity and the measurement accuracy of the key physics objects.
In addition, Particle Flow can largely improve the accuracy of jet energy resolution, since the majority of jet energy is carried by the charged hadrons, whose track momentum are usually measured at a much better accuracy by the tracking system comparing to its cluster energy measured by the calorimeter system.
As the other side of the coin, the software and the reconstruction is vital, and challenge for the Particle Flow oriented design. 
Adequate Particle Flow algorithm is needed to fully exploit the potential of the physics performance. 

A Particle Flow algorithm, Arbor~\cite{Arbormanqi}, has been developed for the CEPC study. 
Arbor has been optimized on a set of reference detector geometries for the CEPC~\cite{CEPCPreCDR}\cite{CEPC_v4Manqi}. 
In this manuscript, we summarize the reconstruction performance at the physics objects and at the Higgs physics benchmarks, based on Geant4~\cite{Geant4Official} simulation. 
The detector geometry is introduced in section~\ref{SGS}.
Section~\ref{ArborPrinciple} briefly summarizes the principle and key performance of the Arbor.
From section~\ref{LeptonID} to section~\ref{JetFlavorTagging}, we demonstrate the reconstruction performance of different physics objects.
Final section~\ref{Summary} is devoted to the conclusion and discussion.

\section{ Reference detector geometry and softwares}
\label{SGS}

To fulfill the CEPC physics requirements, the Particle Flow oriented design is used as the baseline for the CEPC detector design. 
In this manuscripts, most of the results are based on the detector model CEPC v\_1, the benchmark geometry used in the CEPC PreCDR study~\cite{CEPCPreCDR}.
CEPC v\_1 is developed from the ILD detector, the baseline detector of the linear collider studies~\cite{ILC}\cite{CLIC}.
To get adapted to the CEPC collision environments, CEPC v\_1 takes mandatory changes at the Machine Detector Interface (MDI), the forward region, and the Yoke system. 
Comparing to ILC, CEPC requires much short distance between the final focusing magnet (QD0) to the interaction point, which is reduced from 3.5 meters to 1.5 meters. 
The forward region is changed, providing a solid angle coverage of $|cos(\theta)| < 0.995$.
In the original design, the ILD has a total weight of 15k tons, roughly 5 times larger than the LEP detectors. 
The main reason for ILD to be so heavy is its extremely thick return Yoke (3.2 meters in the barrel and 2.6 meters in the endcap). 
Such a heavy yoke is required for the Push-Pull operation scenario, where two detectors are housed in the same experimental Hall and efficient magnetic field shielding is required. 
At CEPC v\_1, the Yoke thickness is reduced by 1 meter for both barrel and endcap and the total weight is reduced by 40\% w.r.t the ILD.

The CEPC v\_1 uses the Time Projection Chamber (TPC) as the main tracker. 
The TPC provides good energy resolution, excellent track reconstruction efficiency and has low material budgets. 
These properties are highly appreciated in the PFA reconstruction. 
The low material budget is important to limit the probability of nuclear interactions and bremsstrahlung before the particle incident on the calorimeter.
In addition, the TPC $dE/dx$ measurement is essential for the charged Kaon identification, see section~\ref{KaonID}.
Using dedicated hardware designs, the TPC is operational at CEPC, where the typical physics event rate at CEPC is roughly 10/1000 Hz at the Higgs/Z pole operation~\cite{TPCFeasibility}.

The TPC in the CEPC v\_1 has a radius of 1.8 meters and a length of 4.7 meters. 
It is divided into 220 radical layers, each has a thickness of 6~mm.
Along the $\phi$ direction, each layer is segmented into 1~mm wide cells. 
In total, the TPC has 10 million readout channels in each endcap. 
Operating in 3.5 Tesla solenoid B-Field, the TPC provides a spatial resolution of 100 $\mu$m in the $R-\phi$ plane and 500 $\mu$m resolution in the $Z$ direction for each tracker hit.
The TPC reaches a standalone momentum resolution of $\delta(1/P_{t}) \sim 10^{-4} GeV^{-1}$.

The CEPC v\_1 is equipped with large-area silicon tracking devices, including the pixel vertex system, the forward tracking system, and the silicon inner/external tracking layers located at the boundary of the TPC. 
Combining the measurements from the silicon tracking system and the TPC, the track momentum resolution could be improved to $\delta(1/P_{t}) \sim 2\times10^{-5} GeV^{-1}$.
In fact, the TPC is mainly responsible for the pattern recognition and track finding, while the silicon tracking devices dominate the momentum measurement.
The silicon pixel vertex system also provides precise impact parameter resolution ($\sim 5\mu m$), which is highly appreciated for the $\tau$ lepton reconstruction and the jet flavor tagging. 

The CEPC v\_1 uses high granular sampling Electromagnetic Calorimeter (ECAL) and Hadronic Calorimeter (HCAL).
The calorimeter is responsible for separating final state particle showers, measuring the neutral particle energy, and providing information for the lepton identification~\cite{FDmanqi}\cite{LICHDan} and charged kaon identification, see section~\ref{KaonID}.
The entire ECAL and HCAL are installed inside the solenoid, providing 3-dimensional spatial position, the energy and the time information for each hit. 
The ECAL is composed of 30 layers of alternating silicon sensor and tungsten absorber.
It has a total absorber thickness of 84~mm. 
Transversely, each sensor layer is segmented into 5~mm by 5~mm cells. 
The HCAL uses Resistive Plate Chamber sensor and Iron absorber. 
It has 48 longitudinal layers, each consists of a 25~mm Iron absorber. 
Transversely, it is segmented into 10~mm by 10~mm cells. 

This calorimeter system provides decent energy measurement for the neutral particles (i.e. roughly $16\%/\sqrt{E/GeV}$ for the photons and $60\%/\sqrt{E/GeV}$ for the neutral hadrons). 
More importantly, it records enormous information of the shower spatial development, ensuring efficient separation between nearby showers and providing essential information for the lepton identification, see section~\ref{LeptonID}. 
In addition, the silicon tungsten ECAL could provide precise time measurement. 
Requesting a cluster level time resolution of 50~ps, the ECAL Time of Flight (ToF) measurement plays a complementary role to the TPC $dE/dx$ measurement, leading to a decent charged Kaon identification performance, see section~\ref{KaonID}.

On top of the CEPC v\_1 geometry, several standalone detector geometries are used to explore the dependence between detector geometry and the objective performances. 
This information is given in corresponding sections. 

All the geometries are implemented via Mokka~\cite{MokkaRef}, the Geant4 simulation package that had been used in the linear collider studies. 
A set of single particle samples and Higgs physics process samples have been used in this manuscript. 
The Higgs physics processes are generated using Whizard~\cite{WhizardRef}. 
The simulated data files are then reconstructed via ilcsoft~\cite{ilcsoftRef} and Arbor. 
The ilcsoft provides functionalities of the data management~\cite{Marlin}\cite{LCIO}, the digitization~\cite{MarlinReco}, the tracking~\cite{Clupatra}, and the flavor tagging.
The Arbor is used as the core PFA algorithm that builds all the reconstructed particles from calorimeter hits and tracks. 
In the next section, we will introduce Arbor.

\section{Arbor}
\label{ArborPrinciple}

Arbor~\cite{Arbormanqi} algorithm is inspired by the simple fact that the particle shower spatial configuration naturally follows a tree configuration.
Arbor is composed of a calorimeter clustering module and a matching module. 
The clustering module reads the calorimeter hits and builds the calorimeter clusters. 
The matching module identifies the calorimeter clusters induced by charged particles (charged clusters), combines these clusters with tracks, and builds charged reconstructed particles. 
The remaining clusters are reconstructed into photons, neutral hadrons, and fragments (mainly from charged clusters). 
The final state particles are therefore reconstructed. 

Arbor clustering module creates oriented connectors between calorimeter hits,
and iterates until the configuration of the connector-hit ensemble follows a tree topology.
The branches hence represent the trajectory of charged shower particles.
The seeds usually correspond to the incident position of the particle at the calorimeter.
Since the separation of the seeds is straightforward, Arbor efficiently separates the particle showers, which is highly appreciated by the Particle Flow principle.

\begin{figure}
\begin{center}
\resizebox{0.45\textwidth}{!}
{
\includegraphics{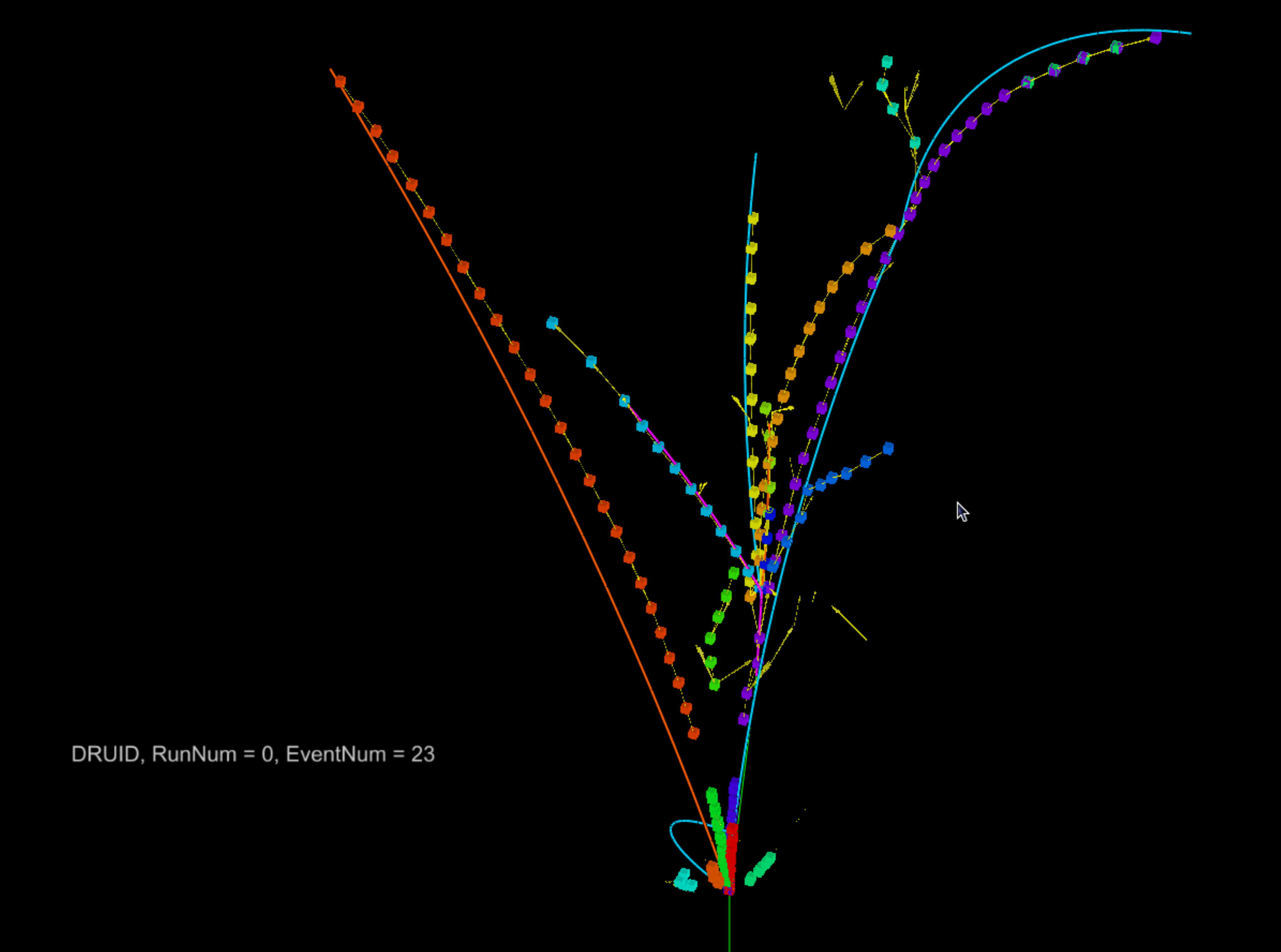}
}
\caption{$K_L$ shower reconstructed by the Arbor algorithm, the branches $-$ the calorimeter hit clusters $-$ are corresponding to the trajectories of charged particles generated in the shower cascade.}
\label{fig:arbor-principle}
\end{center}      
\end{figure}

\begin{figure}
\resizebox{0.5\textwidth}{!}
{
  \includegraphics{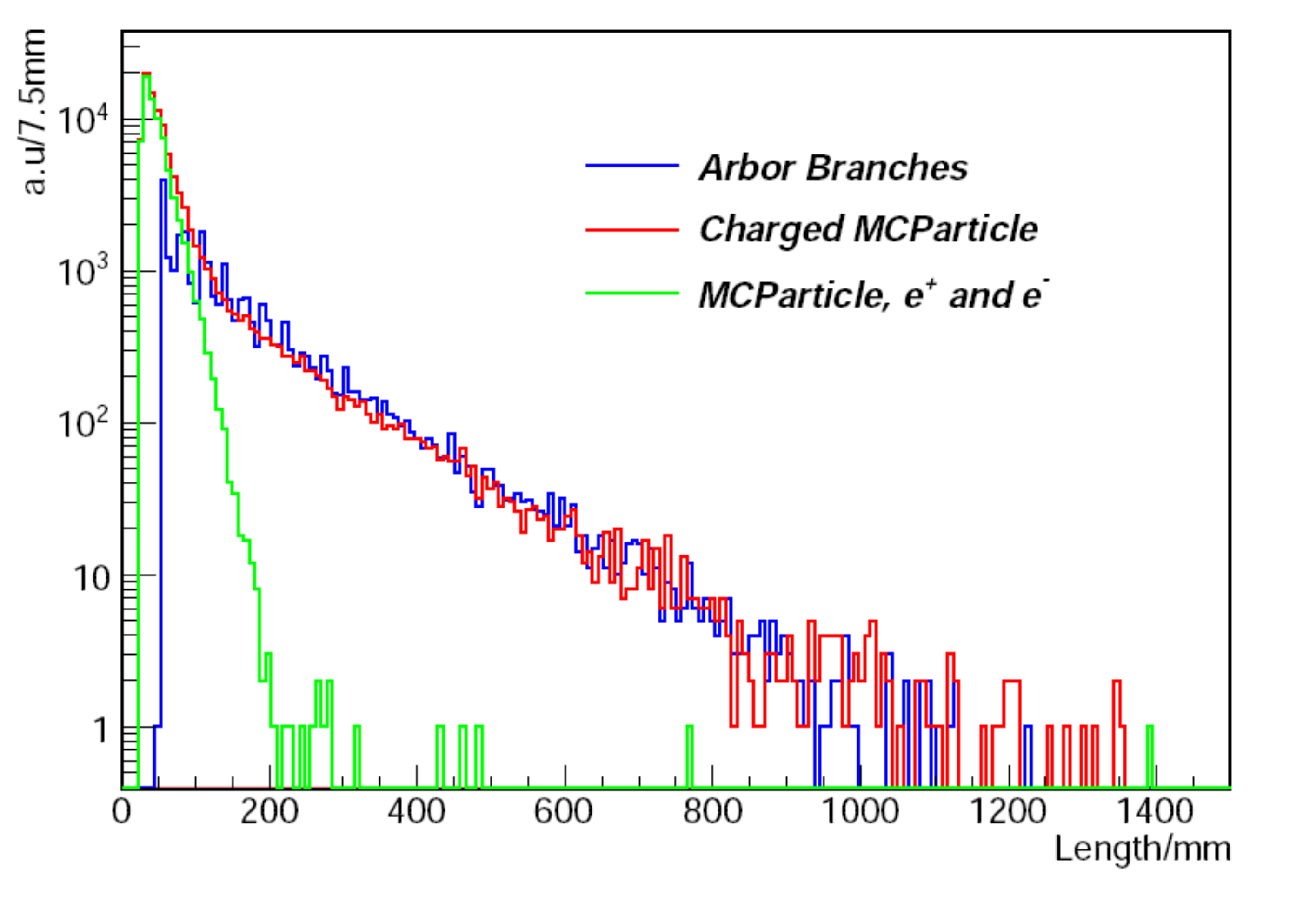}
}
\caption{Proof of Principle:  reconstructed and MC truth particle trajectory length at 40~GeV $\pi$ showers.}
\label{fig:arbor-PoP}      
\end{figure}

Fig.~\ref{fig:arbor-principle} shows a reconstructed calorimeter shower of a 20~GeV $K^{0}_{L}$ particle at the high granularity calorimeter, where the readout density is roughly 1~channel/cm$^3$.
The reconstructed tree branches are demonstrated with different colors.
Therefore the trajectory length of charged shower particle can be reconstructed. 
Fig.~\ref{fig:arbor-PoP} compares the reconstructed trajectory length with MC truth, 
the red distribution is the MC truth level trajectory length of charged particles generated inside 40~GeV $\pi$ showers; 
the green one is corresponding to the trajectory of the electron and the positron generated in the showers; 
while the blue is the trajectory length reconstructed by Arbor. 
Good agreement between the reconstruction and MC truth is found at sufficient trajectory length.

Arbor can also be characterized by the energy collection performance at single neutral particle and the separation performance at bi-particle samples.
Typically, Arbor reaches an energy collection efficiency higher than 99\% for photons with energy higher than 5~GeV.
Higher hit collection efficiency usually leads to a better energy resolution,
however, it usually increases the chance of confusions,
i.e, the wrong clustering of calorimeter hits.
Therefore, an optimized performance depends on the balance of these two effects.

\begin{figure}
\resizebox{0.5\textwidth}{!}
{
  \includegraphics{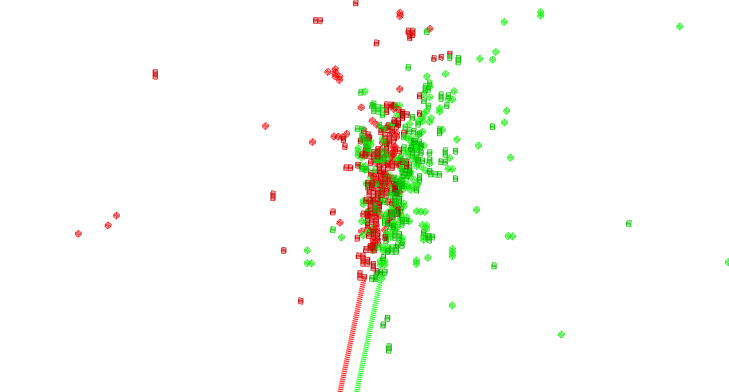}
}
\caption{A reconstructed di-photon event at Si-W ECAL with 1~mm cell size. Each photon has an energy of 5~GeV, and their impact points are separated by 4~mm.}
\label{fig:Di-Photon-Display}      
\end{figure}

\begin{figure}
\resizebox{0.4\textwidth}{!}
{
  \includegraphics{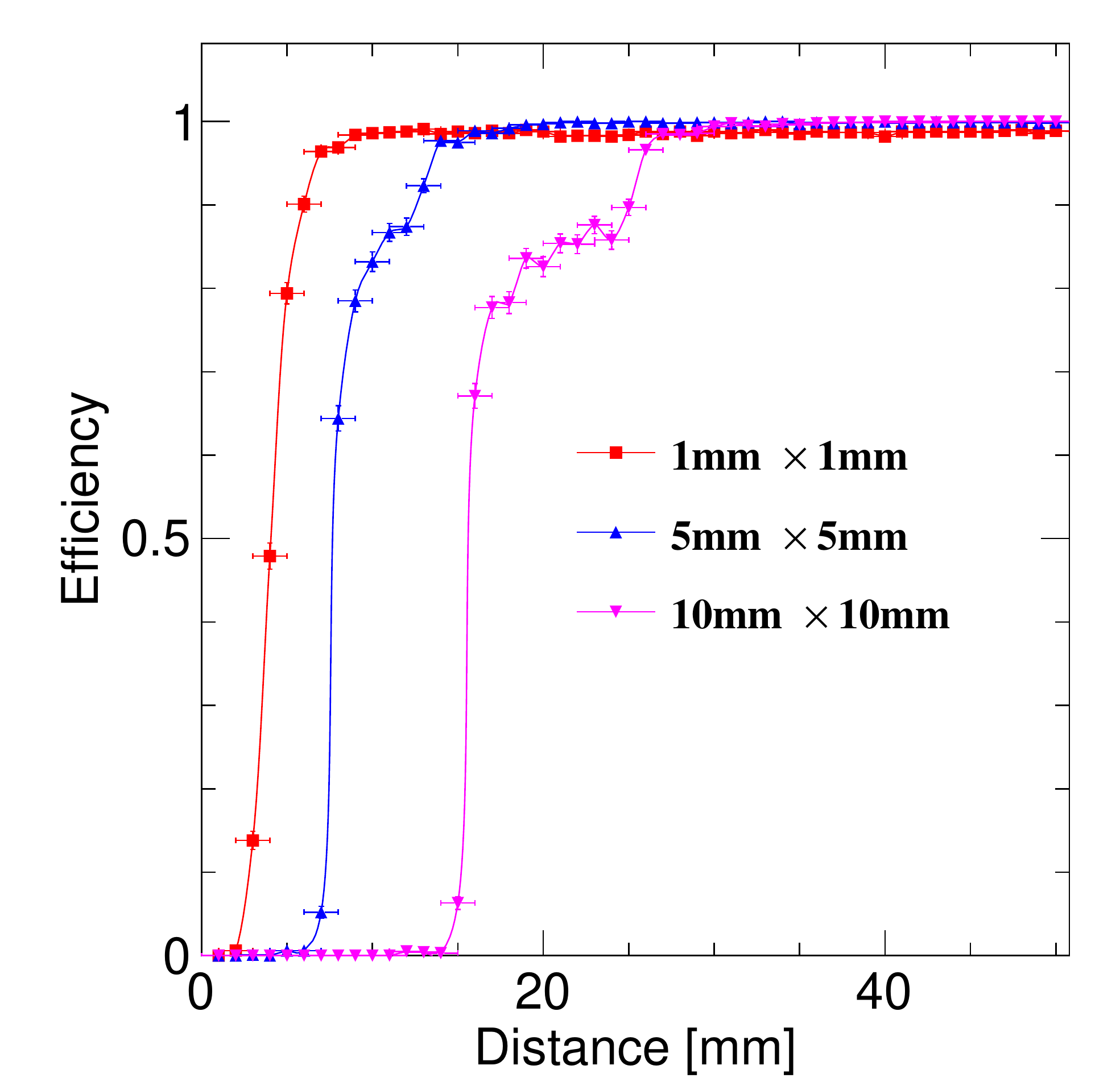}
}
\caption{Reconstruction efficiency of the di-photon events at different ECAL cell sizes. The X-axis represents the distance between photon impact points.}
\label{fig:SepEff}      
\end{figure}

Excellent separation performance is crucial for the jet energy reconstruction, the $\pi^{0}$ reconstruction, and the measurement with $\tau$ final states. 
This performance can be characterized via the reconstruction efficiency of di-photon samples, where two photons with the same energy are shot in parallel at different positions, see Fig.~\ref{fig:Di-Photon-Display}.
According to the distribution of $\pi^{0}$ energy at $Z \to \tau^+\tau^-$ events at CEPC Z pole operation, we set the photon energy to 5~GeV. 

The reconstruction efficiency is defined as the probability of successfully reconstructed two photons with anticipated energy (each candidate is required to have an energy within 1/3 to 2/3 of the total induced energy). 
The efficiency curve naturally exhibits an S-curve dependency on the distance between the photon impact positions, see Fig.~\ref{fig:SepEff}.
The distance at which 50\% of the events are successfully reconstructed is referred to as the critical distance, which depends on the ECAL transverse cell size. 
At the cell size smaller than the Moliere radius, the critical distance is roughly 2 times the cell size, see Table.~\ref{tab:CriticalDis}. 

\begin{table}
\begin{center}
\begin{tabular}{cc}
\hline\noalign{\smallskip}
ECAL cell size & Critical distance for separation \\
\noalign{\smallskip}\hline\noalign{\smallskip}
1~mm & 4~mm \\
5~mm & 9~mm \\
10~mm & 16~mm \\
\noalign{\smallskip}\hline
\end{tabular}
\caption{Arbor critical separation distance at di-photon sample with different ECAL cell size.}
\label{tab:CriticalDis} 
\end{center}
\end{table}

To conclude, Arbor is a geometrical algorithm that reconstructs each shower cluster into a tree topology.
At high granularity calorimeter, Arbor efficiently separates nearby particle showers and reconstructs the shower inner structure.
It maintains a high efficiency in collecting the shower hits/energy, which is appreciated by the energy reconstruction.
The overall performance on different physics object and physics benchmarks will be discussed in details in the following sections.

\section{Leptons}
\label{LeptonID}
The lepton identification is fundamental to the CEPC physics program.
About 7\% Higgs bosons at the CEPC are generated with a pair of leptons.
Those events are the golden signal for the Higgs recoil analysis, which is the anchor for the absolute Higgs measurements at the electron-positron Higgs factory.
A significant fraction of the Higgs boson decays, directly or via cascade, into final states with leptons. 
In addition, a significant fraction of $H\to bb/cc$ events generate leptons in their jet fragmentation cascade, thus a good lepton identification performance improves flavor tagging performance.
The lepton identification is also crucial for the EW measurements. 

\begin{figure}[h!]
\begin{center}
\resizebox{0.4\textwidth}{!}
{
\includegraphics{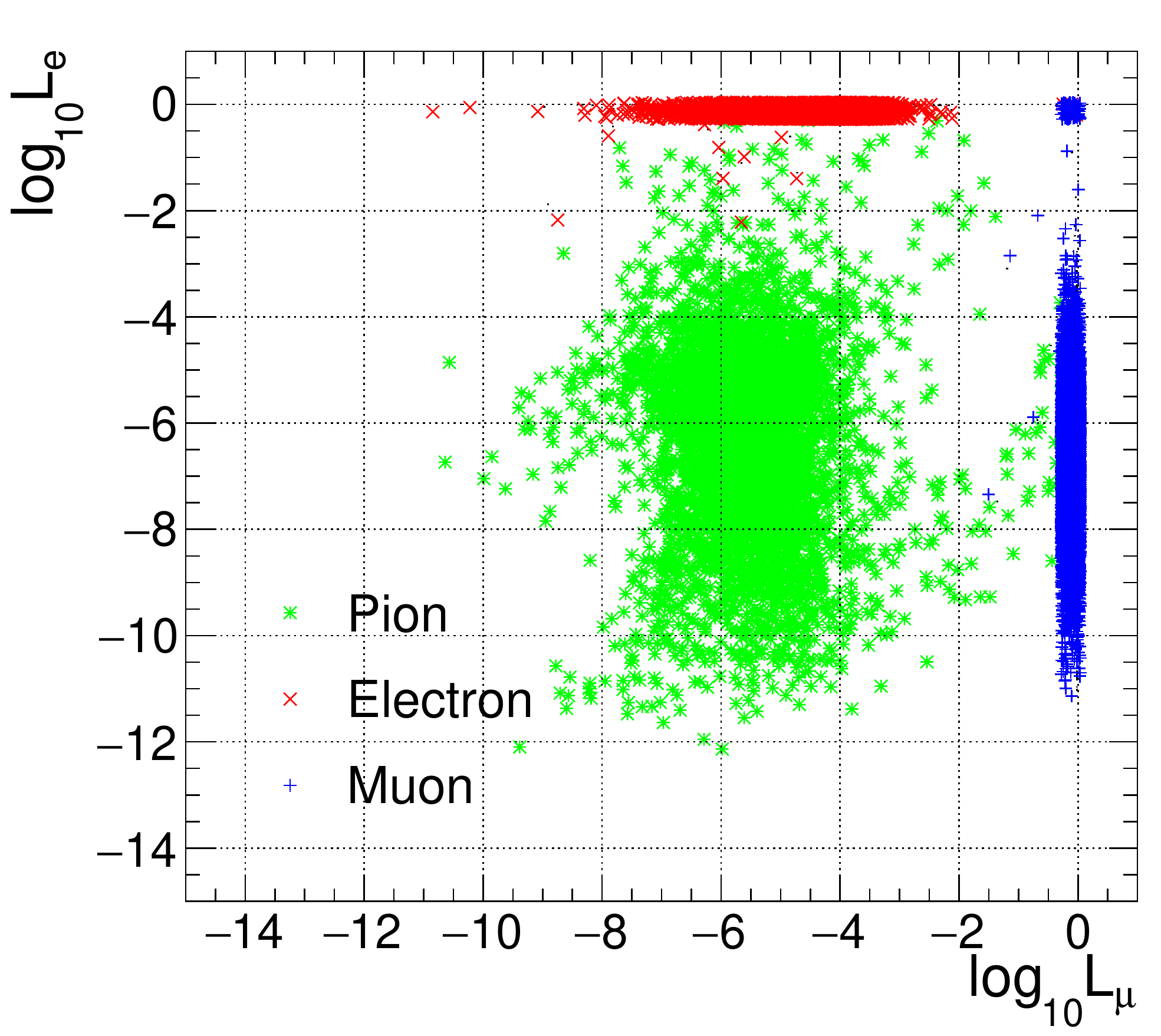}
}
\caption{Lepton likelihood of electron, muon and pion calculated by LICH (using final state particle reconstructed by Arbor).}
\label{fig:leptonlikelihood}
\end{center}
\end{figure}

\begin{figure}
\begin{center}
\resizebox{0.5\textwidth}{!}
{
\includegraphics{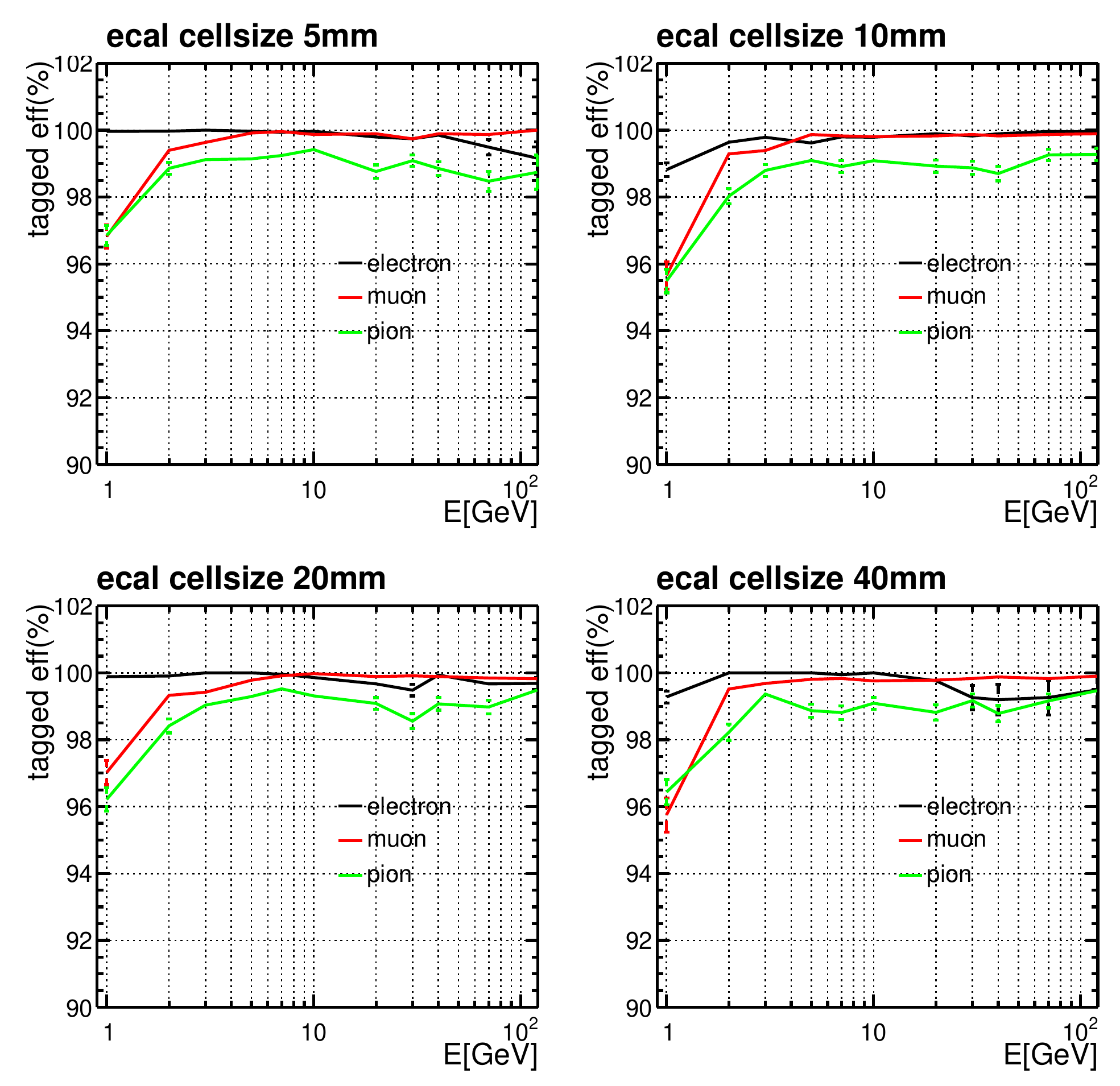}
}
\caption{Efficiencies of $\mu^{\pm}$ (blue), $e^{\pm}$ (red) and $\pi^{\pm}$ (green) identifications at different calorimeter granularity.}
\label{fig:performance-lepton}
\end{center}
\end{figure}

The PFA oriented detector, especially the high granularity calorimeter system, provides enormous information for the lepton identification.
A dedicated lepton identification algorithm, LICH~\cite{LICHDan}, has been developed for the detectors using high granularity calorimeter.
For each reconstructed charged particle, LICH extracts more than 20 observables from the associated track and calorimeter cluster. 
These observables include the track $dE/dx$ measurement, the shower fractal dimension~\cite{FDmanqi} that describes the global shower compactness, the shower longitudinal profiles, and the distances in between the track and calorimeter cluster. 
Using the Gradient Boost Decision Tree method at the TMVA toolkit~\cite{ROOTTMVA}, LICH then calculates the electron and muon likelihood for the charged particle. 
Fig.~\ref{fig:leptonlikelihood} shows the likelihood distribution of 40~GeV electron, muon and pion samples, where clear separation is observed. 


At the CEPC v\_1 geometry, for isolated charged particles with energy larger than 2~GeV, LICH achieves a lepton identification efficiency better than 99.5\%.
The accumulated misidentification rate of hadrons to leptons is smaller than 1\%.
This misidentification is mainly caused by the irreducible background such as pion decays and highly electro-magnetic like pion clusters (via the $\pi^{0}$ generated from the pion-nuclear interactions).
The performance of LICH has been scanned over a large range of the granularity for both ECAL and HCAL, while the performance is stable for particles with energy larger than 2 GeV, see Fig.~\ref{fig:performance-lepton}.

This performance is significantly better than the experiments at the LHC and the LEP~\cite{LeptonIDLEP}\cite{LeptonIDLHC}.
In the physics event, the lepton identification performance is limited by the separation power of the particle detector.
To evaluate this impact, we studied the efficiency of successfully identified two prompt leptons at the $l^{+}l^{-}H$ event.
This analysis shows at 10~mm ECAL cell size, the reconstruction efficiency reaches 97-98\%, for $e^{+}e^{-}H$ and $\mu^{+}\mu^{-}H$ events respectively~\cite{LICHDan}.
This efficiency degrades at larger ECAL cell size.
Taken into account the detector acceptance, we conclude that less than 0.5\% of the prompt leptons in the $l^{+}l^{-}H$ events will potentially be misidentified due to the limited separation power at the CEPC v\_1 geometry.

\section{Charged kaons}
\label{KaonID}

Successful identification of the charged kaons is crucial for the flavor physics and is appreciated in the jet flavor and jet charge measurements~\cite{RomanFrancoisKaon}. 
A clear $\pi-K$ separation is the key for the charged kaon identification. 
According to the Bethe-Bloch equation, the $dE/dx$ of the charged pions is larger than that of kaons by roughly 10\% at the same momentum in the relativistic energy range at the CEPC Z pole operation.
In another word, an efficient $\pi-K$ separation can be achieved if the $dE/dx$ can be measured to a relative accuracy better than 5\%.

\begin{figure}[h!]
\begin{center}
\resizebox{0.4\textwidth}{!}
{
\includegraphics{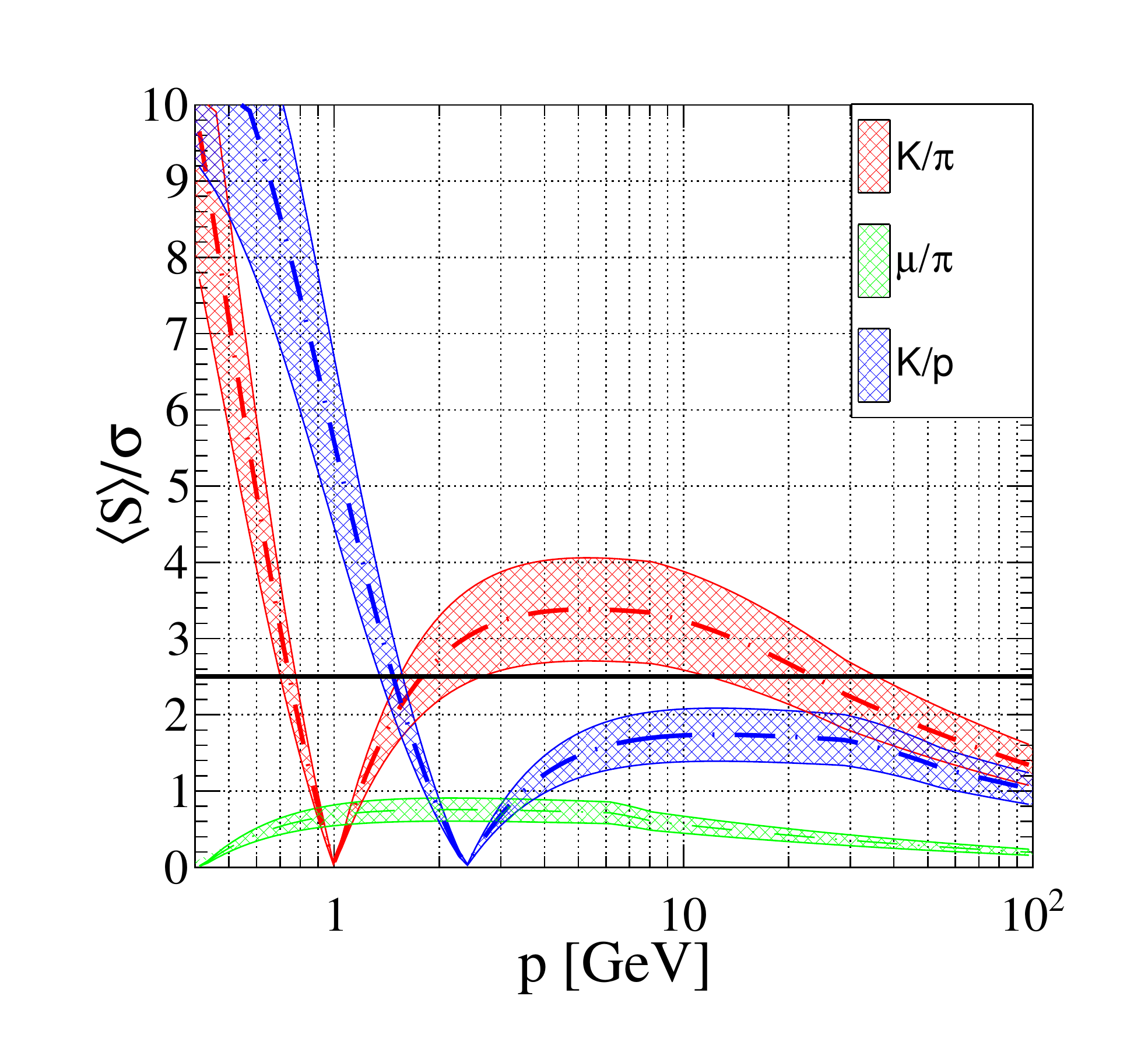}
}
\caption{ $dE/dx$ separation at the CEPC v\_1 detector. The upper boundary is corresponding to the MC truth, the lower boundary includes a 50\% degradation (conservative scenario), the middle curve is corresponding to 20\% degradation (objective scenario).}
\label{fig:dedx}
\end{center}
\end{figure}

The large TPC main tracker at the CEPC v\_1 provides the $dE/dx$ measurement. 
At the MC truth level, the Geant4 simulation predicts a $3.9 \sigma$ $\pi$-$K$ separation and $1.5 \sigma$ $K-proton$ separation at the inclusive $Z \to q \bar{q}$ samples at 91.2 GeV center of mass energy~\cite{KaonIDSoren} (Integrated over track momentum range of 2-20 GeV). 
A survey of the existing experiments shows that, with respect to the MC truth, the achieved $dE/dx$ measurements degrade by 15 - 50\%. 
which is caused by the intrinsic energy resolution, the inhomogeneity, the stability of devices, the occupancy,  {\it etc}.
The 50\% degrading is used as a conservative estimation of the $dE/dx$ measurement at the CEPC. 
Fig.~\ref{fig:dedx} shows the anticipated separation performance between different charged particles at the CEPC v\_1 TPC. 
The upper band boundaries are corresponding to the MC truth prediction, while the lower boundaries are corresponding to this conservative estimation. 
Integrated over the momentum interval of 2-20 GeV, a $2.6 \sigma$ $\pi$-$K$ separation is anticipated in the conservative estimation. 

\begin{figure}[h!]
\begin{center}
\resizebox{0.4\textwidth}{!}
{
\includegraphics{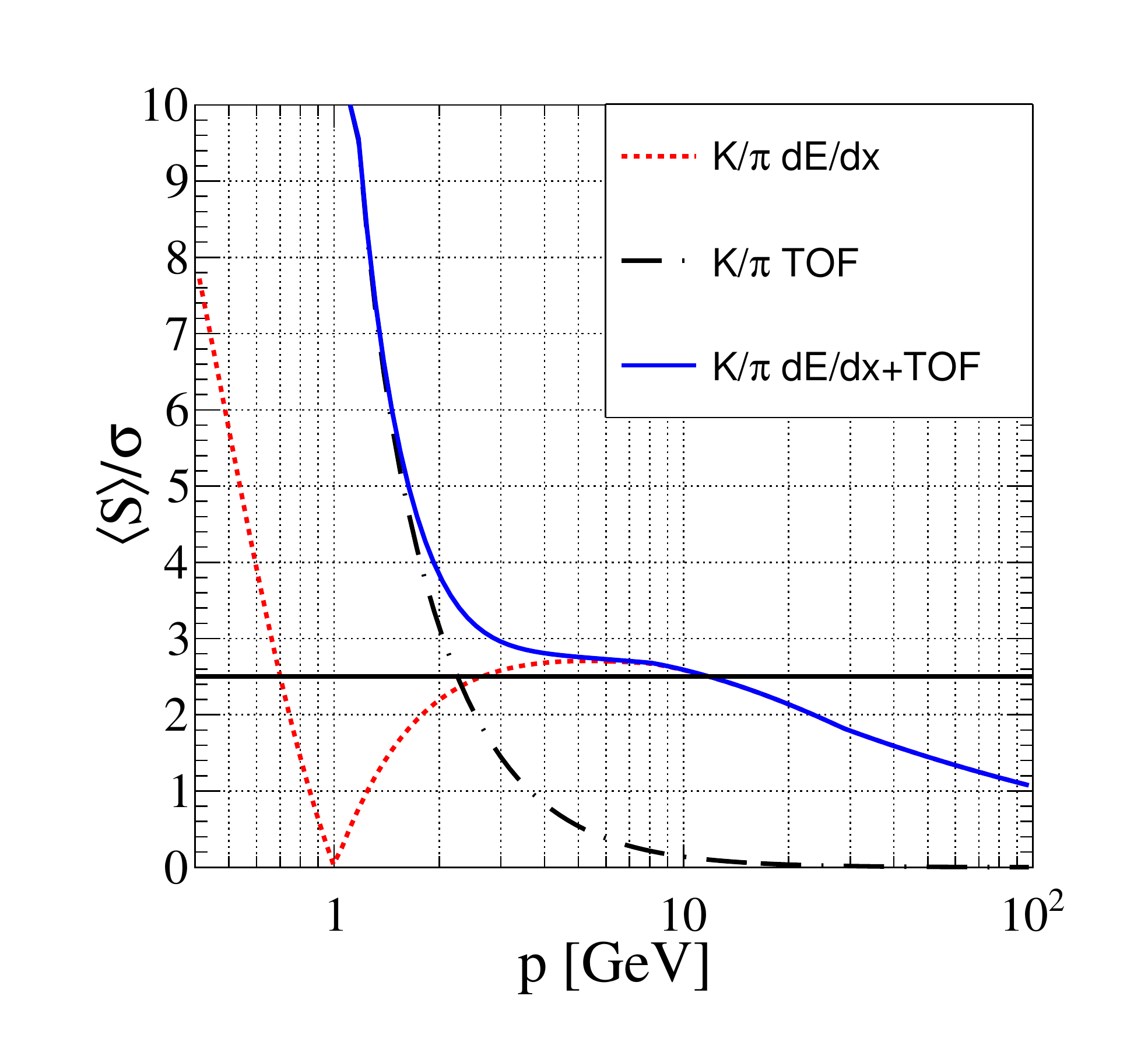}
}
\resizebox{0.4\textwidth}{!}
{
\includegraphics{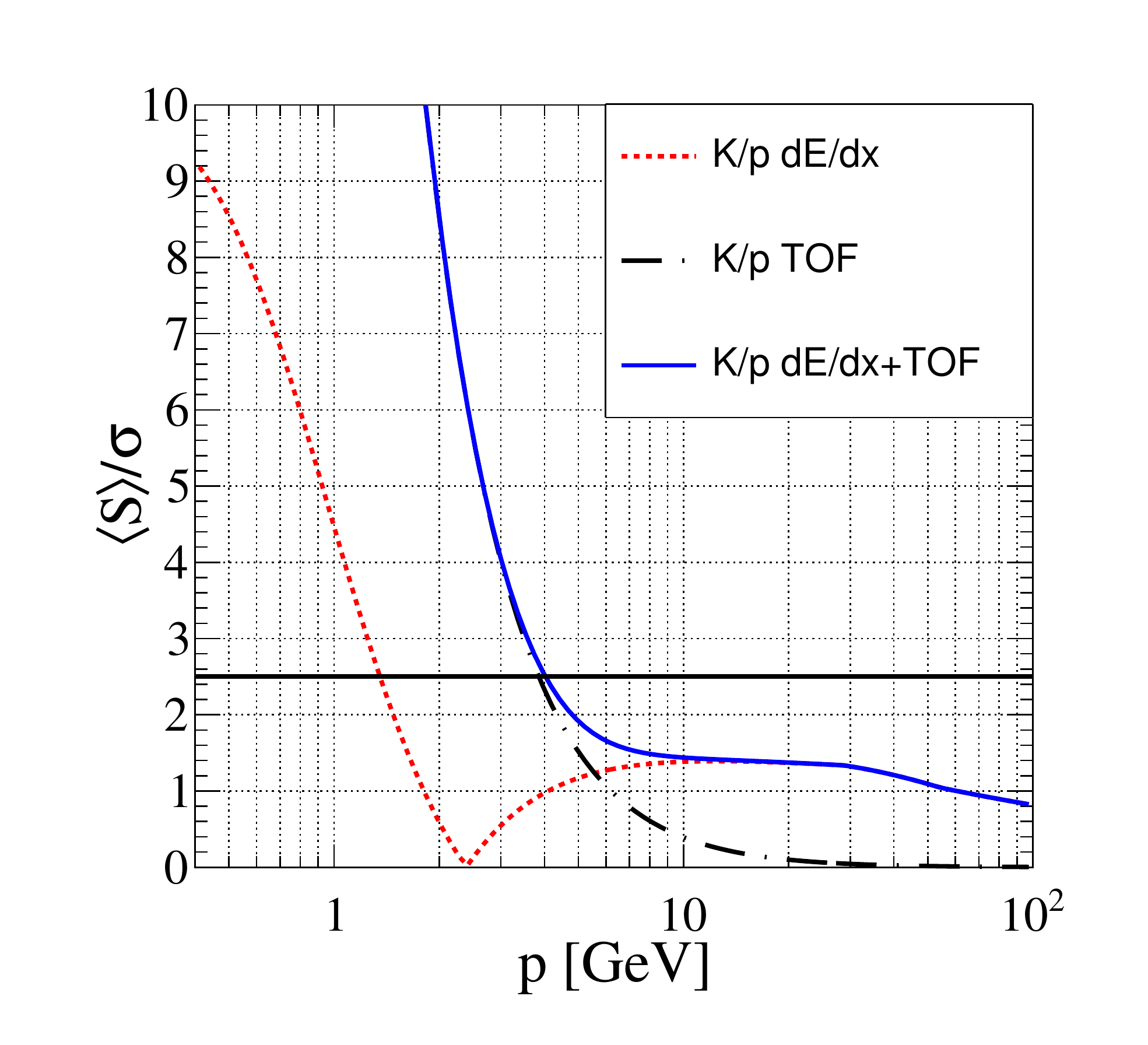}
}
\caption{The separation between $\pi-Kaon$ (upper plot) and $Kaon-Proton$ (lower plot), using the conservative estimation of the TPC $dE/dx$ and the ECAL ToF measurement with 50~ps time resolution at the cluster level.}
\label{fig:dedx2}
\end{center}
\end{figure}

The $dE/dx$ difference between the pions and the kaons vanishes at 1~GeV track momentum.
To cover this low momentum range, a Time of Flight (ToF) measurement with an accuracy of 50~ps (at cluster level) is proposed.
According to the recent progress of high granularity calorimeters, this ToF information could be measured by the ECAL~\cite{CMS-HGC}\cite{ATLAS-HPTD}\cite{Christophe-Timing}.
This ToF measurement is crucial for the $K$-$p$ separation, see Fig.~\ref{fig:dedx2}. 
Using both ToF and $dE/dx$ information, at inclusive $Z\to q\bar{q}$ sample at 91.2 GeV center of mass energy, 
a kaon identification reaches an efficiency/purity of 91\%/94\% in the conservative scenario at the CEPC v\_1 geometry.
If the $dE/dx$ measurement achieves an objective scenario that the degrading with respect to the MC truth is controlled to be 20\%, 
the identification performance could be improved to an efficiency/purity of 97\%/97\%, which is only 2\% degraded from the MC truth prediction. 

To conclude, a decent kaon identification performance could be achieved using the TPC $dE/dx$ measurement and the ECAL ToF measurement. 
The TPC hardware design is encouraged to achieve a $dE/dx$ resolution that degrades less than 20\% with respect to the MC truth prediction. 
Benchmarked with tracks at $Z\to q\bar{q}$ events, the $dE/dx$ resolution should be measured to a precision better than 3.6\%.
The ECAL ToF measurement is recommended to achieve a time resolution of 50~ps at the cluster level.

\section{Photons}
\label{PhotonReco}

Successful photon reconstruction is crucial for the jet energy reconstruction, the $Br(H \to \gamma \gamma)$ measurement, and the physics with $\tau$ leptons.
In this study, we benchmark the overall photon reconstruction using the Higgs mass resolution with $H \to \gamma \gamma$ event.

The photon reconstruction is sensitive to the tracker material and the calorimeter geometry defects, such as the cracks between the ECAL modules, staves, and the dead zone between the ECAL barrel and endcaps.
To quantify their impact, a simplified, defect-free ECAL geometry is implemented.
The benchmark Higgs invariant mass distributions are analyzed for both simplified and realistic geometry (the CEPC v\_1). 

\begin{figure}[h!]
\resizebox{0.4\textwidth}{!}
{
\includegraphics{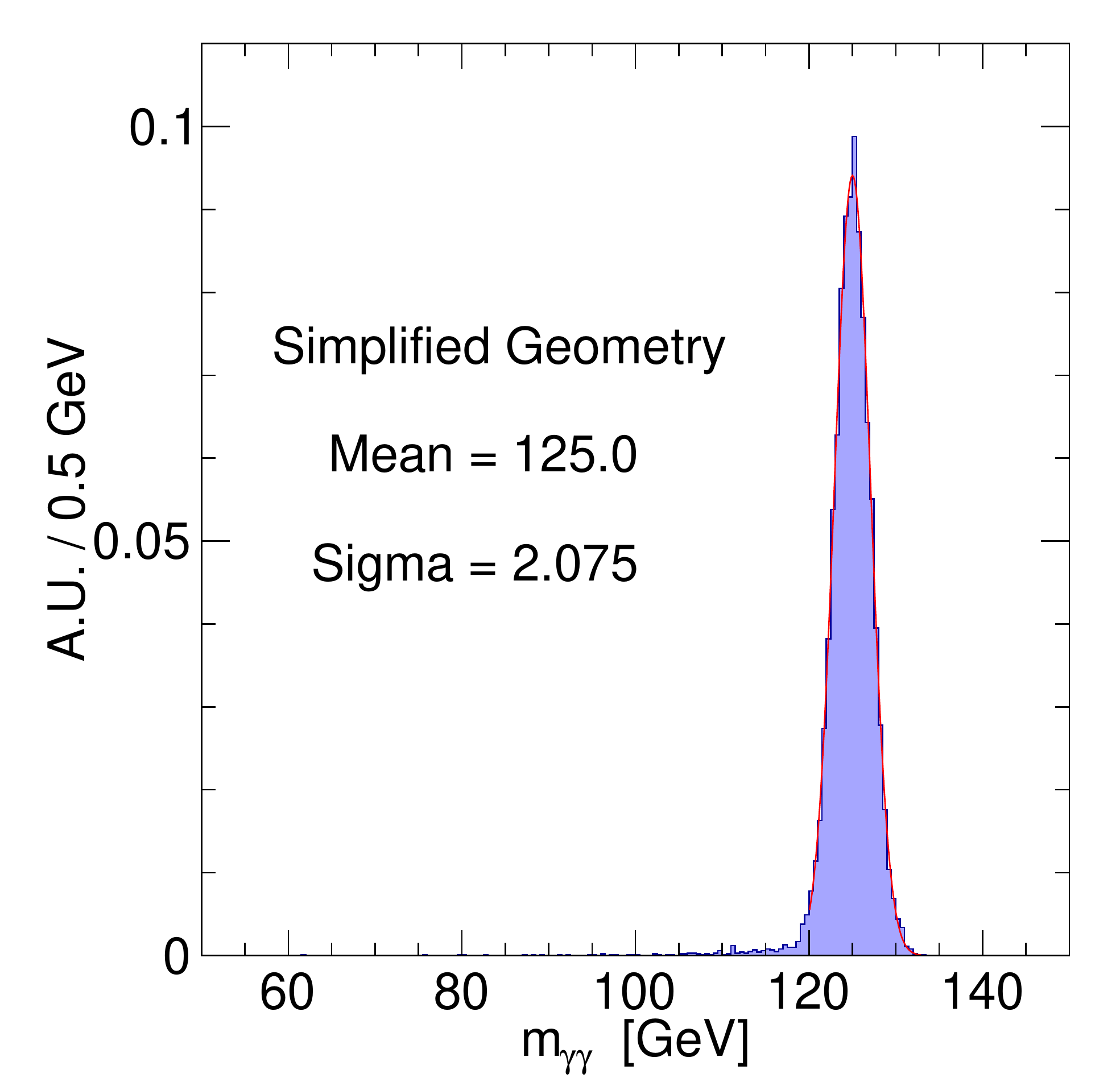}
}
\caption{The reconstructed Higgs invariant mass of $H \to \gamma \gamma$ events at the simplified detector geometry (without any gap and defects in the ECAL, and has no tracker). 10k events are reconstructed and the distribution is normalized to unit area. }
\label{fig:performance-diphoton}
\end{figure}

This simplified geometry uses cylindrical barrel layer and its endcaps are directly attached to the barrel, forming a closed cylinder.
No tracker geometry is implemented in this simplified geometry. 
Fig.~\ref{fig:performance-diphoton} shows the Higgs boson invariant mass reconstructed from $Br(H \to \gamma \gamma)$ signal at this simplified geometry.
A relative mass resolution of 1.7\% is achieved, which agrees with the intrinsic electromagnetic energy resolution measured at the CALICE Si-W ECAL prototype test beam experiments~\cite{CALICEReview}.

\begin{figure}[h!]
\resizebox{0.4\textwidth}{!}
{
\includegraphics{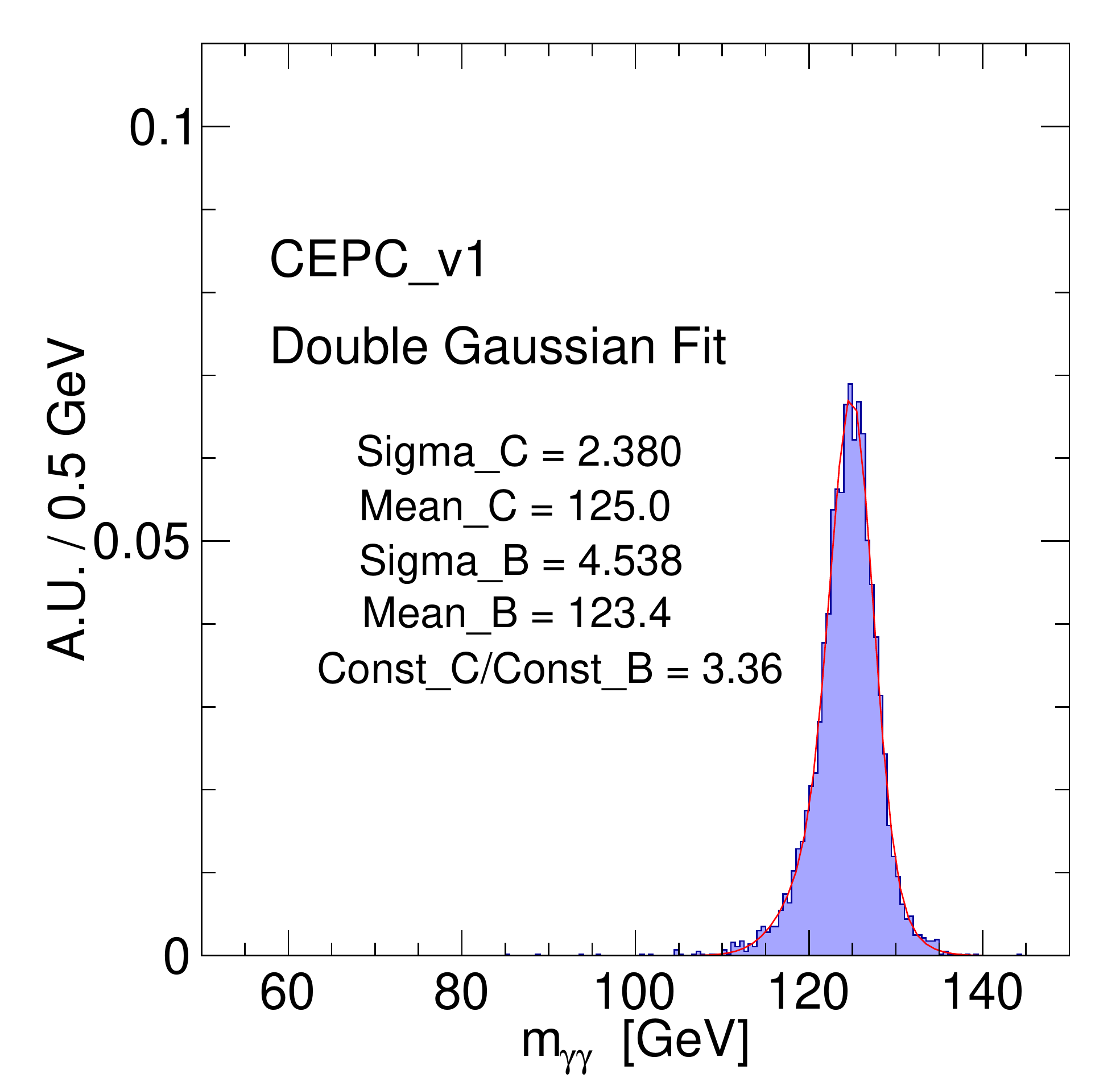}
}
\caption{The reconstructed Higgs invariant mass of $H \to \gamma \gamma$ events at the CEPC v\_1 detector geometry. 6k events, normalized to unit area. }
\label{fig:performance-diphoton_v1}
\end{figure}

Comparing to the simplified geometry, the relative resolution of the Higgs mass at CEPC v\_1 degrades by almost a factor of two, and the mean value of the mass peak is shifted to 121~GeV. 
A preliminary geometry based correction algorithm has been developed, which scales the energy of EM clusters located at the geometry cracks.
After applying this correction algorithm, the Higgs boson invariant mass distribution at CEPC v\_1 is shown in Fig.~\ref{fig:performance-diphoton_v1}.
This distribution could be fit to a core Gaussian center and a wider Gaussian with a lower mean value. 
The core gaussian exhibits a mass resolution of 1.9\%, while the low-mass wider gaussian is caused by the fact that the correction algorithm is only optimized.
The average mass resolution (taking weighted average of both Gaussian) is then 2.3\%.
The latter can be improved with much dedicated correction algorithm. 

In terms of photon reconstruction efficiency, the CEPC v\_1 detector is sensitive to photons with energy larger than 10 MeV, the efficiency saturates to 100\% for photon energy larger than 1 GeV~\cite{ManqiArborPresentation}.
Proportional to the material before the calorimeter, roughly 7\% of the photons at CEPC v\_1 convert into $e^+e^-$ pairs or even start an electromagnetic shower before reaching the calorimeter.
Thanks to the lepton identification performance and the large solid angle coverage, the majority of these converted photons could be identified.

To summarize, our simulation predicts the Higgs mass resolution at two-photon final state reaches 1.6-2.1\% level at the CEPC.
This result is consistent with the CALICE prototype test beam result.
The reconstruction of converted photons and the correction of the geometry defects at any realistic detector geometry is vital for the photon reconstruction. 

\section{Taus}
\label{TauReco}

The $\tau$ lepton is an extremely intriguing physics object.
As the heaviest lepton in the SM, $\tau$ has a large Yukawa coupling to the Higgs boson, leading to a significant $Br(H\to\tau^{+}\tau^{-})$.
The $\sigma(HX)\times Br(H\to\tau^{+}\tau^{-})$ is expected to be measured better than 1\% relative accuracy at the CEPC~\cite{DanThesis}.
Measuring the $\tau$ polarization at the $Z$ pole leads to a precise determination of $\sin^2\theta^{eff}_{W}$~\cite{LeptonIDLEP}. 
Also, the measurements via spectral functions of $\tau$ hadronic decays are very compelling at the CEPC~\cite{TauPhysicsALEPH}.

The $\tau$ lepton has various different decay modes, and the successful $\tau$ lepton identification is highly non-trivial.
In the CEPC studies, we classify the events with final state $\tau$ leptons into two classes and develop the identification algorithms accordingly.

The first class is the leptonic events, whose final states contain no jet, for example:

\begin{itemize}
\item[]1, $e^+e^- \to ZH, Z\to ll$ or $\nu\bar{\nu}, H \to \tau^+\tau^-$ events;

\vspace*{0.2cm}

\item[]2, $e^+e^- \to ZZ \to ll/\nu\bar{\nu} + \tau^+\tau^-$ events;

\vspace*{0.2cm}

\item[]3, $WW$ events with $l\nu \tau\nu$ final states;

\vspace*{0.2cm}

\item[]4, $Z \to\tau^+\tau^-$ events at CEPC $Z$ pole operation.

\end{itemize}

A successful identification of these events based mostly on the reconstruction of photons, charged particles, and the track impact parameters.

The second class is the hadronic events with jets in their final states, for instance:

\begin{itemize}
\item[]1, $ZH\to q\bar{q}\tau^+\tau^-$

\vspace*{0.2cm}

\item[]2, $ZZ \to q\bar{q} \tau^+\tau^-$

\vspace*{0.2cm}

\item[]3, $WW \to q\bar{q} \tau\nu$

\end{itemize}

Finding the $\tau$ candidate in the hadronic events depends on the isolation conditions, the multiplicities, the visible mass of $\tau$ candidates, and the track impact parameters.

A full simulation analysis of $g(H\tau^{+}\tau^{-})$ measurement includes both classes and is performed at~\cite{DanThesis}. 
The first class is represented by the $Br(H\to\tau^+\tau^-)$ measurement at $\mu^+\mu^- H$ events.
The inclusive SM background is efficiently subtracted by requesting the proper multiplicity of photons, charged particles and the restriction on the invariant/recoil mass of the $\mu^+\mu^-$ system.
Thanks to the PFA oriented design and reconstruction, the final event selection reduced the inclusive SM background by nearly six orders of magnitudes, while preserves a signal efficiency of 93\%.
The leading remaining background is the irreducible Higgs background (i.e. $H\to WW^{*}, ZZ^{*} \to \tau^+\tau^-\nu\tilde{\nu}$).
A relative accuracy of 2.7\% is achieved for the signal strength measurement in the $\mu^+\mu^- H$ channel.

\begin{figure}[h!]
\begin{center}
\resizebox{0.4\textwidth}{!}
{
\includegraphics{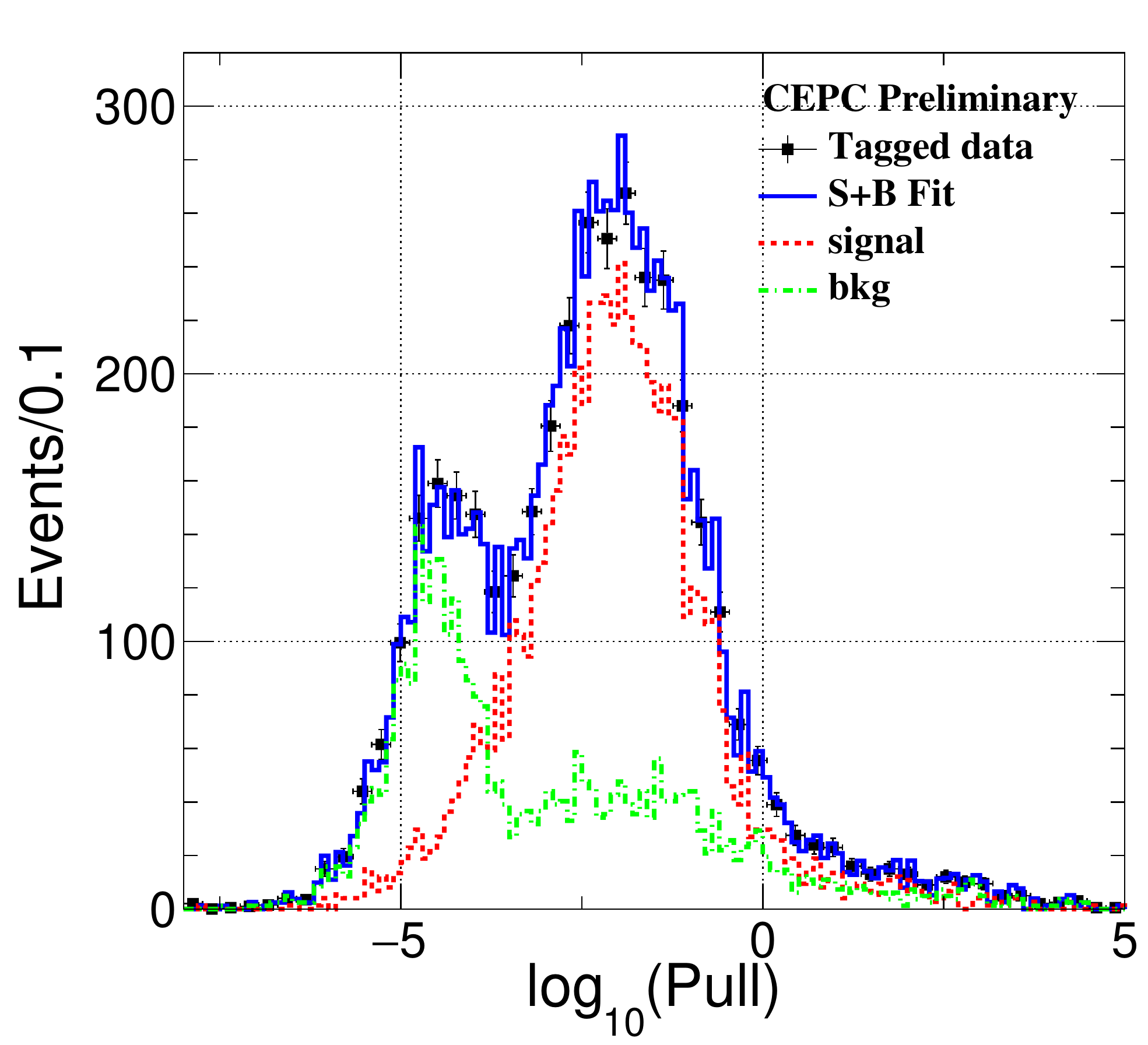}
}
\resizebox{0.4\textwidth}{!}
{
\includegraphics{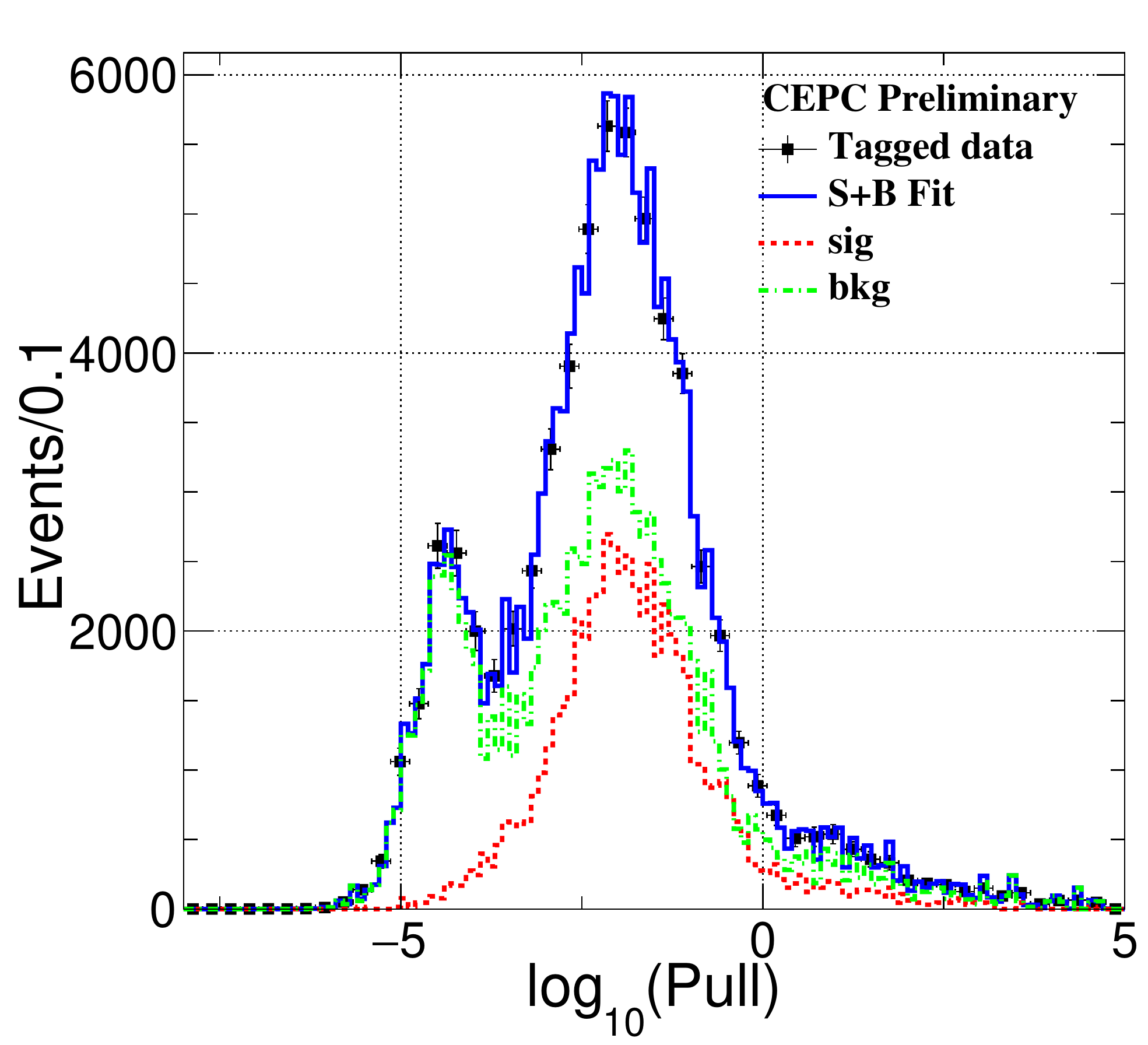}
}
\caption{The pull of impact parameters at $Br(H \to \tau^+\tau^-)$ measurement via $\mu^+\mu^-H$ (up) and $q\bar{q}H$ channel(down).}
\label{fig:tautau}
\end{center}
\end{figure}

The second class includes $q\bar{q}H, H\to\tau^+\tau^-$ events.
A double size cone-based $\tau$ finding algorithm is developed.
For each individual track, two cones with different sizes are formed. 
A $\tau$ candidate is identified once the multiplicities, the mass, {\it etc} at each cone satisfy certain constraints. 
These cone parameters are optimized.
In short, by requesting two $\tau$ candidates with opposite charge,
the signal efficiency is 57\% and the background could be suppressed by three orders of magnitude.

Giving the significant c$\tau$ of the $\tau$ lepton (89 $\mu m$) and the precise vertex system at CEPC v\_1, the signal and background could be further separated using the track impact parameter $D_0$ and $Z_0$. 
For each track, we define a pull parameter as $((D_0/mm)^2 + (Z_0/mm)^2$. 
Fig.~\ref{fig:tautau} shows the sum of the pull of the leading track for each tau candidate for both signal and backgrounds (after above-mentioned event selection), where the signal is clearly separated from the background for both $\mu^+\mu^- H$ and $q\bar{q}H$ channels.
Applying a template fit to the pull parameter, a relative accuracy of 2.1\% and 1.0\% for the signal strength measurements can be achieved for the $\mu^+\mu^- H$ and $q\bar{q}H$ channels respectively. 

To conclude, the $\tau$ reconstruction at the CEPC uses different algorithms for the leptonic and hadronic events.
In both cases, the $\tau$ events identification relies strongly on a successful reconstruction of the photons, charged hadrons, and leptons, 
which, is secured by separation performance of Arbor with current CEPC baseline detector geometry.
Meanwhile, a precise reconstruction of the impact parameters plays an important role in the identification of events with $\tau$ final states.

It should be reminded that the requirements of $\tau$ physics are more demanding than the $g(H\tau^+\tau^-)$ measurements.
The former requests a successful reconstruction of the number of $\pi^0$ generated in the $\tau$ decay cascade, 
making strong requirements on the separation power of ECAL and on the ECAL energy/geometry acceptances.

\section{Jet}
\label{JetReco}

The jet is fundamental for the CEPC physics program. 
About 90\% of the SM Higgs boson decays into final states with jets (70\% directly to di-jet final states; and roughly 20\% via decay cascade from the $ZZ^{*}, WW^{*}$),
while 70\% of W and Z bosons decay into di-jet final states. 
Roughly 60\% of the jet energy is carried by the charged particles, and the Particle Flow could improve significantly the precision of jet energy measurement with respect to the calorimeter based reconstruction.

In the Particle Flow reconstruction, the jet candidates are constructed from the reconstructed final state particles via the jet clustering algorithms.
The ambiguity from the jet clustering is significant and usually dominants the uncertainty, especially for these events with more than two final state jets such as the measurement of $g(Hb\bar{b})$, $g(Hc\bar{c})$, and $g(Hgg)$ via $ZH\to4 jet$ events. 

To characterize the jet reconstruction performance, a two-stage evaluation has been applied at the CEPC studies. 
The first stage is the Boson Mass Resolution (BMR) analysis designed to avoid the complexity induced by the jet clustering. 
The second is the individual jet response analysis, which requests the jet clustering. 

The Boson Mass Resolution analysis is applied to physics events with two final state jets decayed mostly from one intermediate gauge boson, including
\begin{itemize}
\item [] 1, $\nu\tilde{\nu} q\bar{q}$ events via the $ZZ$ intermediate state;

\vspace*{0.2cm}

\item [] 2, $l\nu q\bar{q}$ events via mostly $WW$ intermediate state;

\vspace*{0.2cm}

\item [] 3, $\nu\tilde{\nu} H$ events with $H \to b\bar{b}, c\bar{c},~\mbox{or}~gg$.

\end{itemize}

In these processes, besides the jet final state particles, the other particles are either invisible or could be easily identified. 
The invariant mass of all the boson final state particles can be reconstructed.
Therefore, disentangled from the jet clustering algorithm, the BMR evaluates the jet reconstruction. 
Meanwhile, the BMR shows immediately how these massive gauge bosons can be separated at jet final state. 

Using the jet clustering and matching algorithms, the jet response is also analyzed at each individual jet. 
The overall response includes the detector resolution, the ambiguous induced by the jet clustering and the mismatching. 
These effects are physics process dependent and a complete analysis is beyond the scope of this manuscript. 
In this paper, this analysis is limited to individual jet reconstruction performance at $\nu\tilde{\nu} q\bar{q}$ process.

\begin{figure}[h!]
\begin{center}
\resizebox{0.4\textwidth}{!}
{
\includegraphics{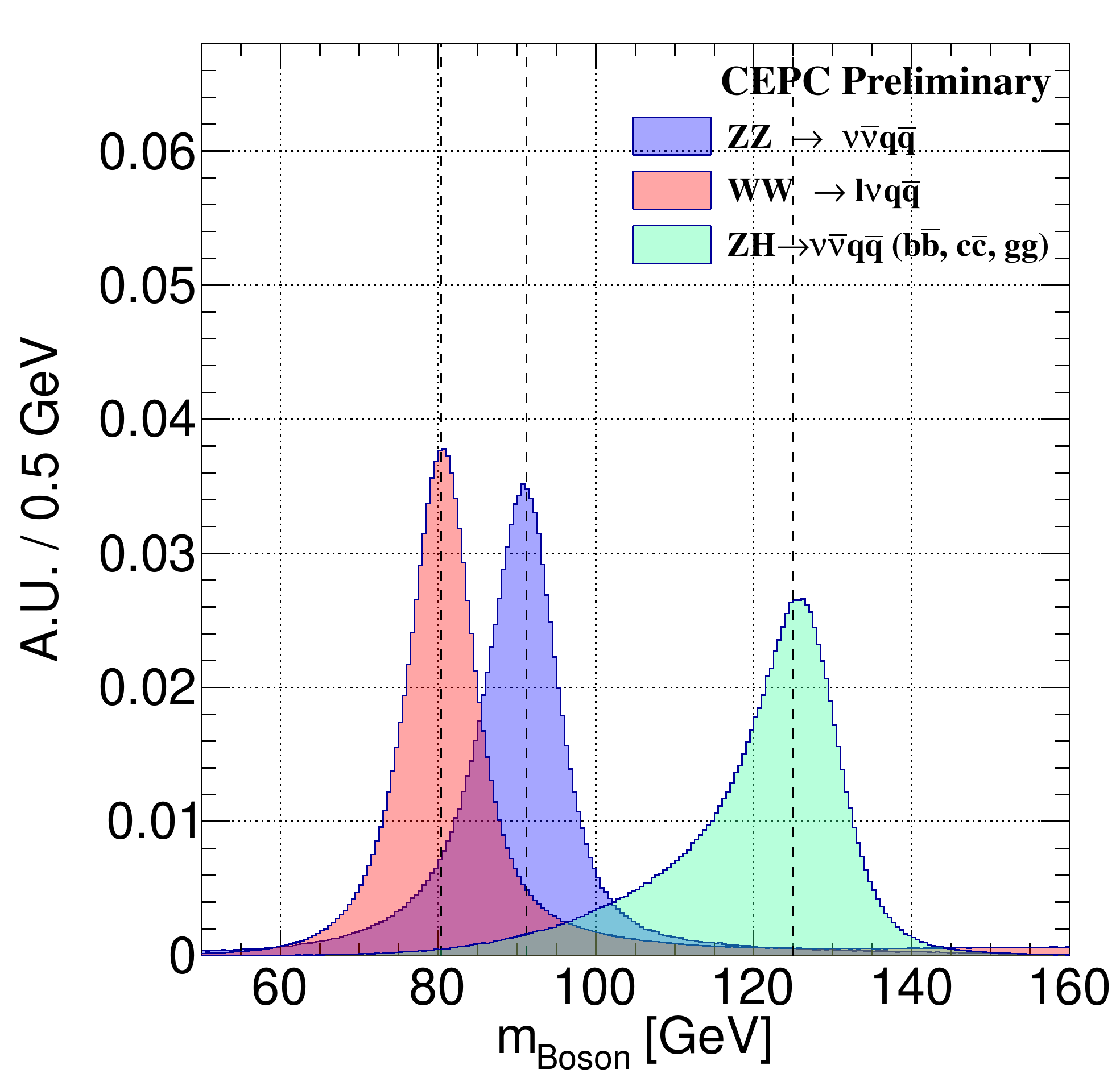}
}
\caption{Reconstructed boson masses of the inclusive $l\nu q\bar{q}$ (red), $\nu\tilde{\nu} q\bar{q}$ (blue) and $\nu \tilde{\nu} H, H\to jj$ samples (green).}
\label{fig:jet-HZW-1}
\end{center}
\end{figure}

Corresponding to 5~$ab^{-1}$ integrated luminosity at the CEPC, 
we simulate 1.8 millions $\nu\tilde{\nu} q\bar{q}$, 11 millions $l\nu q\bar{q}$ and 170 thousands $\nu\tilde{\nu} H, H\to jj$ events at the CEPC v\_1 geometry.
All these samples are reconstructed with Arbor. 
Fig.~\ref{fig:jet-HZW-1} shows the inclusive reconstructed boson mass distributions normalized to unit area. 
These distributions are well separated, each exhibits a peak at the expected boson mass. 
These mass distributions are all asymmetric for different reasons.
At the low mass side, the green distribution, corresponding to $\nu\tilde{\nu} H, H\to jj$ events, has a long tail.
This tail is mainly stemmed from the neutrinos generated in the heavy jets fragments (most of the $H\to jj$ events are $H\to b\bar{b}$ events ).
The heavy jet components are also responsible for the low mass tail in the other two distributions. 
Because W boson hardly decays into b-jets, the low mass tail of $l \nu q\bar{q}$ sample is much less significant. 
The Breit-Wigner width of massive gauge bosons and the phase space effects also contribute to the long tails at the $l \nu q\bar{q}$ and the $\nu\tilde{\nu} q\bar{q}$ samples.
The high mass tail induced by ISR photon(s) is observed in each distribution. 

\begin{figure}[h!]
\begin{center}
\resizebox{0.4\textwidth}{!}
{
\includegraphics{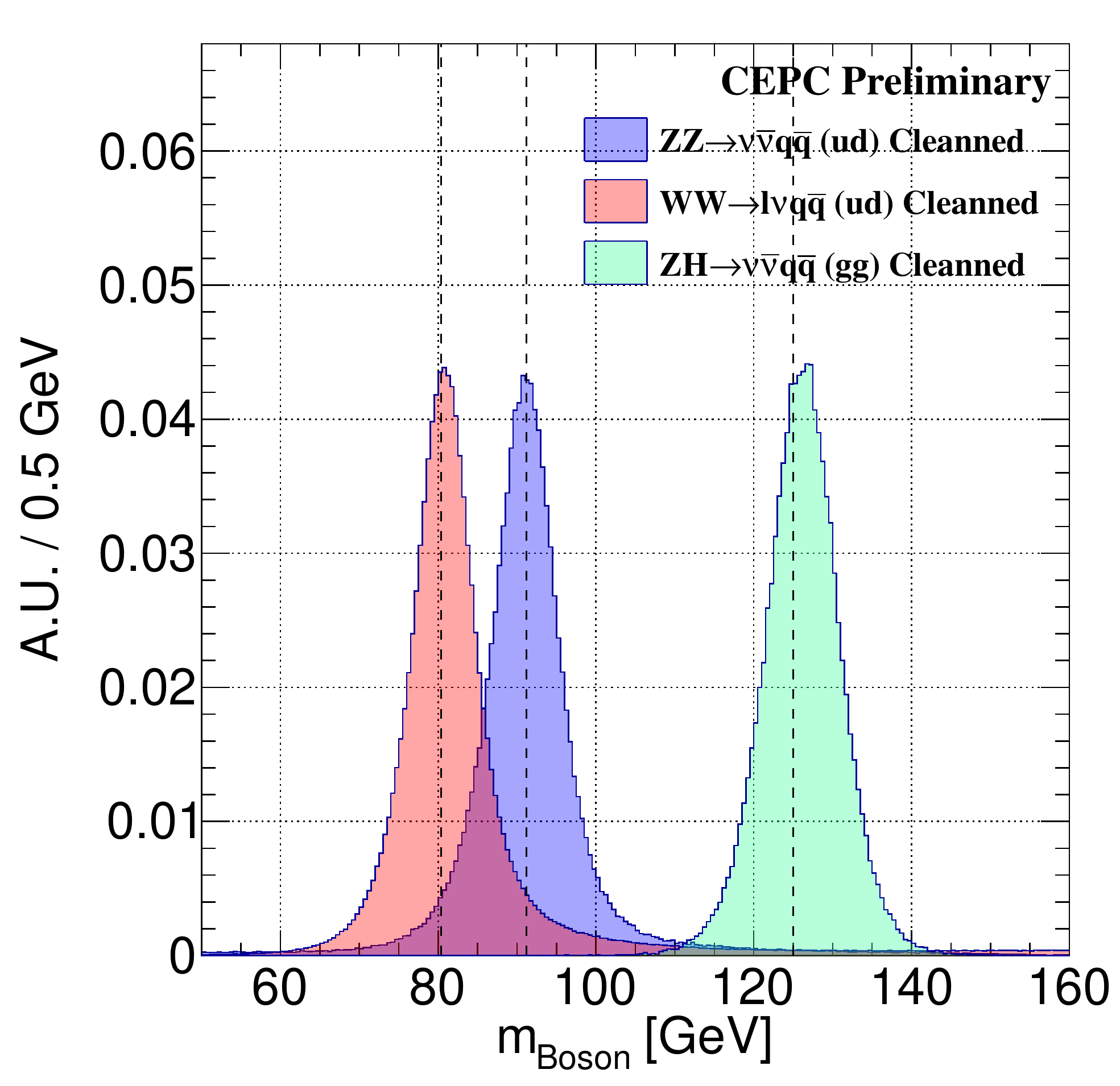}
}
\caption{Reconstructed boson masses of the cleaned $\nu\tilde{\nu}q\bar{q}$ (red), $l\nu q\bar{q}$ (blue) and $\nu \tilde{\nu} H, H\to gg$ (green). The requirements are described in the main text.}
\label{fig:jet-HZW-2}
\end{center}
\end{figure}

To decouple the detector response from these physics effects, a standard event selection is designed: 
\begin{itemize}
\item [] 1, the jets are generated from light flavor quarks ($u$, $d$) or gluons. 

\vspace*{0.2cm}

\item [] 2, the partons should have a significant angle from the beam pipe: $|cos(\theta)| < 0.85$.

\vspace*{0.2cm}

\item [] 3, there is no energetic visible final state ISR photon: the accumulated scalar transverse momentum of the ISR photons should be smaller than 1~GeV. 

\vspace*{0.2cm}

\item [] 4, there is no energetic jet neutrino: the accumulated scalar transverse momentum of the jet neutrinos should be smaller than 1~GeV. 

\end{itemize}
This event selection clearly leads to much narrow boson mass distribution and much better separation, see Fig.~\ref{fig:jet-HZW-2}.

After this event selection, the mass distributions are much symmetric. 
The Higgs boson mass could be simply fit to a Gaussian, while the other two distributions include the non-negligible intrinsic widths.  
The efficiency of this event selection depends on the decay branching ratio (condition 1), differential cross section (condition 2), the radiation behavior (condition 3) and jet fragmentation (condition 4). 
 As in the $\nu \tilde{\nu} H, H\to gg$ sample, this event selection has an overall efficiency of 65\% (75\%/94\%/94\% for the 2nd/3rd/4th condition, respectively).
The relative mass resolution of the Higgs mass is then 3.8\%, providing a quantitative reference for the BMR. 

It should be remarked that both lepton identification and jet flavor tagging information are available from current reconstruction. 
Combing these information enhances the distinguishing power on different physics processes. 

The calibration process plays an important role in measuring the jet energy.  
Technically, Arbor was calibrated via two steps, the single particle level calibration, and the data-driven calibration.
The single particle calibration is to figure out the global ECAL/HCAL calibration constants according to the comparison between the reconstructed neutral particle energy and the truth. 
The ECAL calibration constant is derived from photon samples while the HCAL calibration constant at $K^{0}_{L}$ samples. 
Due to the Particle Flow double counting, i.e. the fragments of charged particle showers are misidentified as neutral particles, the single particle calibration leads to typically 1\% overestimation on the boson mass. 
The data-driven calibration is to scale all the reconstructed boson masses according to the W mass peak exhibited in the $l\nu q\bar{q}$ events, the leading physics processes of the above three.  
This simple calibration simultaneously scales the three boson mass peak positions to the expected positions. 
To fully appreciate the enormous productivity of massive bosons at the CEPC, sophisticated calibration methods must be developed and validated for the real experiments, i.e. control and corrections of differential dependences, in-situ calibrations, detector homogeneity monitoring and control, {\it etc}. 

The reconstruction performance of individual jet is explored via the same $\nu \tilde{\nu} q\bar{q}$ sample. 
Using ee-anti-kt algorithm (a.k.a Durham algorithm~\cite{FastJet}), all the reconstructed particles are forced into two jets (recojets). 
The same jet-clustering algorithm is applied to the visible final state particles at the MC truth level, forming the generator level jets (genjets).
Using a matching algorithm that minimizes the angular difference, the jet reconstruction performance is characterized by the difference between the 4-momentum of the initial quarks, the genjets, and the recojets.
The difference between the quarks and the genjets is mainly coming from the fragmentation and the jet clustering processes, 
while the difference between the genjets and the recojets is induced by the jet clustering, matching, and the detector response. 
A dedicated analysis shows that, even at this simple di-jet process, the uncertainty induced by the jet clustering and matching can be as significant as those from the detector response~\cite{JERNote_Peizhu}. 

These two reconstructed jets are classified into leading/sub-leading jets according to their energy. 
The relative energy difference between genjet and recojet is then fit with a double-sided crystal ball function. 
The exponential tails are mainly induced by the jet clustering algorithm, the matching performance, and the detector acceptance. 
The Gaussian core then describes the detector resolution, therefore we define its mean value as the Jet Energy Scale (JES) and its relative width as the Jet Energy Resolution (JER).

\begin{figure}[h!]
\begin{center}
\resizebox{0.4\textwidth}{!}
{
\includegraphics{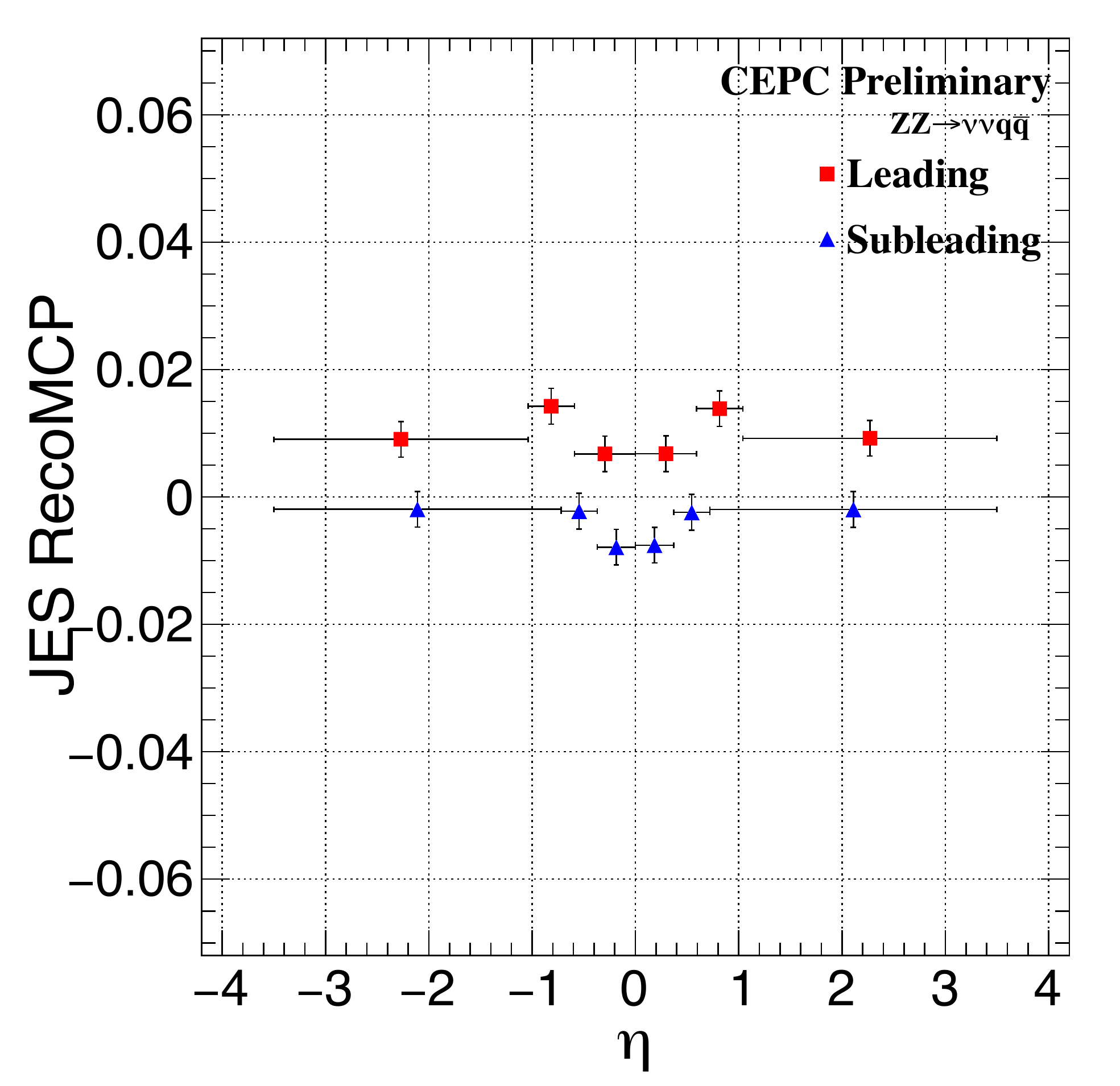}
}
\resizebox{0.4\textwidth}{!}
{
\includegraphics{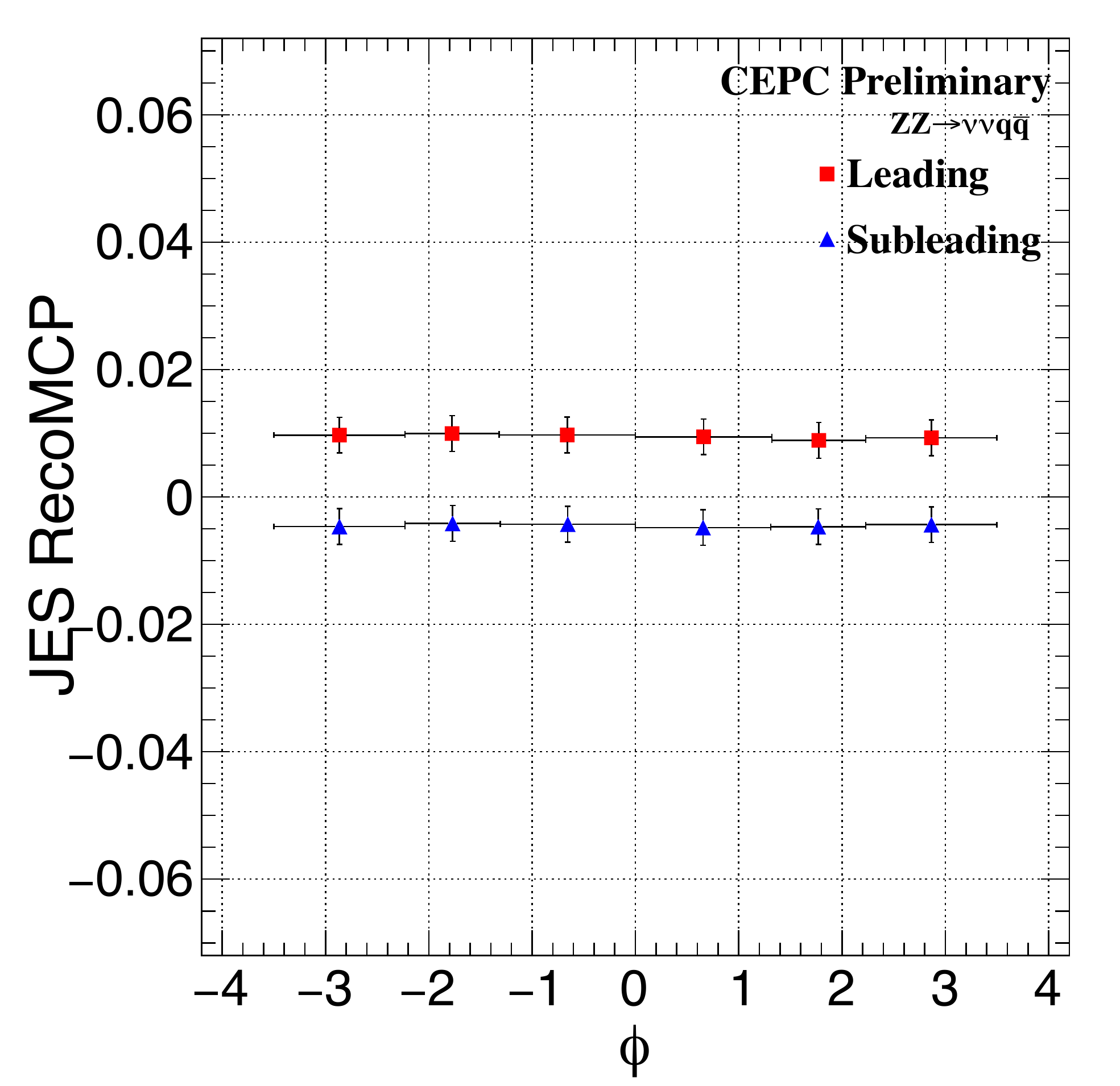}
}
\caption{Jet energy scale at different jet directions.}
\label{fig:jet-energy-scale}
\end{center}
\end{figure}

Fig.~\ref{fig:jet-energy-scale} shows the JES at different jet directions.
The JES is flat along the azimuth angle.
Along the polar angle, the JES increases significantly for the leading jets in the overlap part between the endcap and the barrel.
The JES is also larger in the endcap than in the barrel.
These patterns are correlated with the Particle Flow confusions, especially the artificial splitting of the charged clusters.
Not surprisingly, the leading jets have a systematically higher JES comparing to the sub-leading one.
Without any corrections, the entire amplitude of the JES is controlled to 1\% level, which is significantly better than that of LHC even after the correction~\cite{CMSJet}.

\begin{figure}[h!]
\begin{center}
\resizebox{0.4\textwidth}{!}
{
\includegraphics{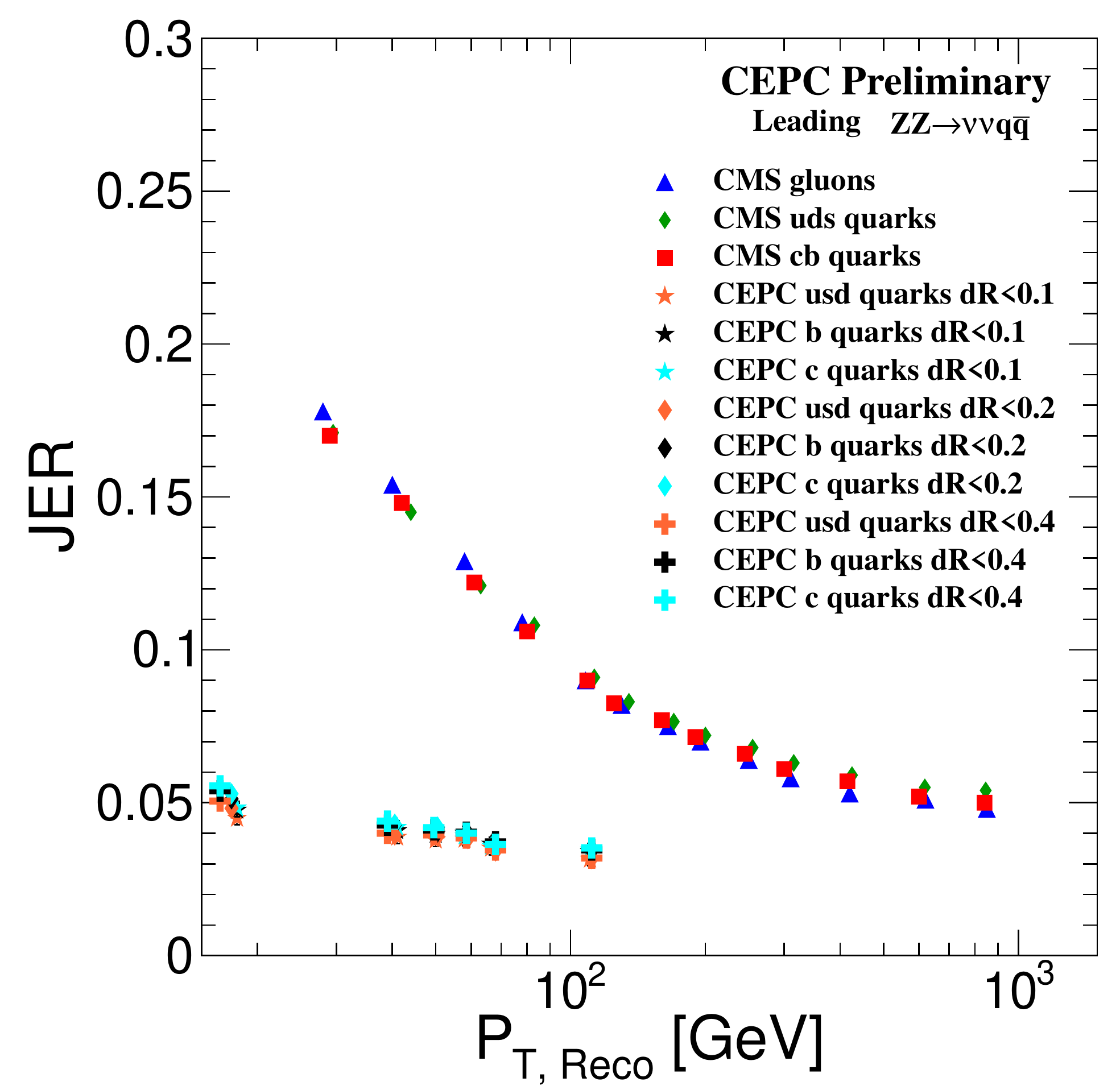}
}
\resizebox{0.4\textwidth}{!}
{
\includegraphics{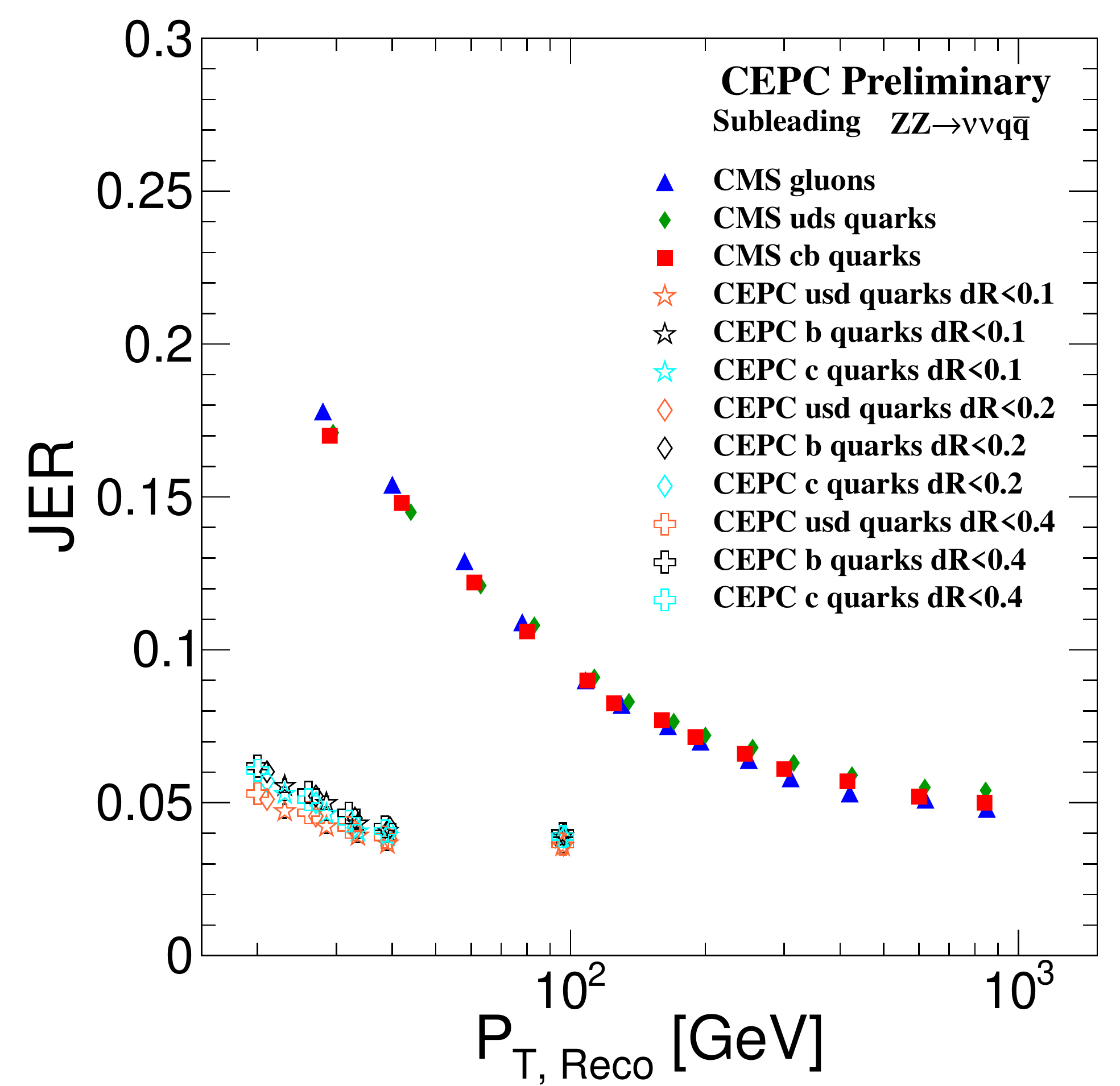}
}
\caption{
  The jet energy resolution for leading (upper) and sub-leading jets (lower), as a function of the jet transverse momenta. The performance at the CMS~\cite{CMSJet} has been overlapped for comparison.
  }
\label{fig:jet-comparison}
\end{center}
\end{figure}

The jet energy resolution (JER) at different jet transverse momenta is displayed in Fig.~\ref{fig:jet-comparison}.
The overall JER takes a value between 6\% (at $P_t < 20$ GeV) to 3\% (at $P_t > 100$ GeV).
The leading jets usually has a slightly better JER comparing to the sub-leading ones.
Taking the performance of the CMS detector as a reference,
the JER at the CEPC reference detector is 2-4 times better at the same $P_t$ range~\cite{CMSJet}.

To conclude, the jet energy response has been analyzed at the BMR level and at the individual jet level.
For physics events with only two jets, the boson mass could be measured to a relative accuracy better than 4\% at CEPC v\_1 using a standard event selection.
This resolution ensures significant separation between the $W$ boson, the $Z$ boson, and the Higgs boson.
At individual jets, the JES is controlled to 1\% level and the JER of 3\% to 6\%, both are significantly better than the LHC detector performances. 
This superior performance is based on the clean electron-positron collision environment, the PFA oriented detector design and reconstruction. 
It is highly appreciated for the CEPC physics program, i.e. the measurements of W boson mass at the CEPC Higgs operation. 
It should also be emphasized that the jet-clustering algorithm has a strong and even dominant impact on the physics measurements with multiple jets in the final states. 

\section{Jet Flavor Tagging}
\label{JetFlavorTagging}

Identification of the jet flavor is essentially for the measurement of the Higgs couplings ($g(Hb\bar{b}), g(Hc\bar{c}), g(Hgg)$) and the EW observables at the CEPC.
During the jet fragmentation cascade, the heavy flavor quarks ($b$ and $c$) are mostly fragmented into heavy hadrons (i.e. $B^{0}$, $B^{\pm}$, $B_{s}$, $D^{0}$, $D^{\pm}$, {\it etc}). 
Those heavy hadrons have a typical $c\tau$ of a few hundred micrometers. 
Therefore, the reconstruction of the secondary vertex is crucial for the flavor tagging. 
The information of jet mass, vertex mass, number of leptons, {\it etc}, are also frequently used in flavor tagging. 

\begin{figure}[h!]
\begin{center}
\resizebox{0.4\textwidth}{!}
{
\includegraphics{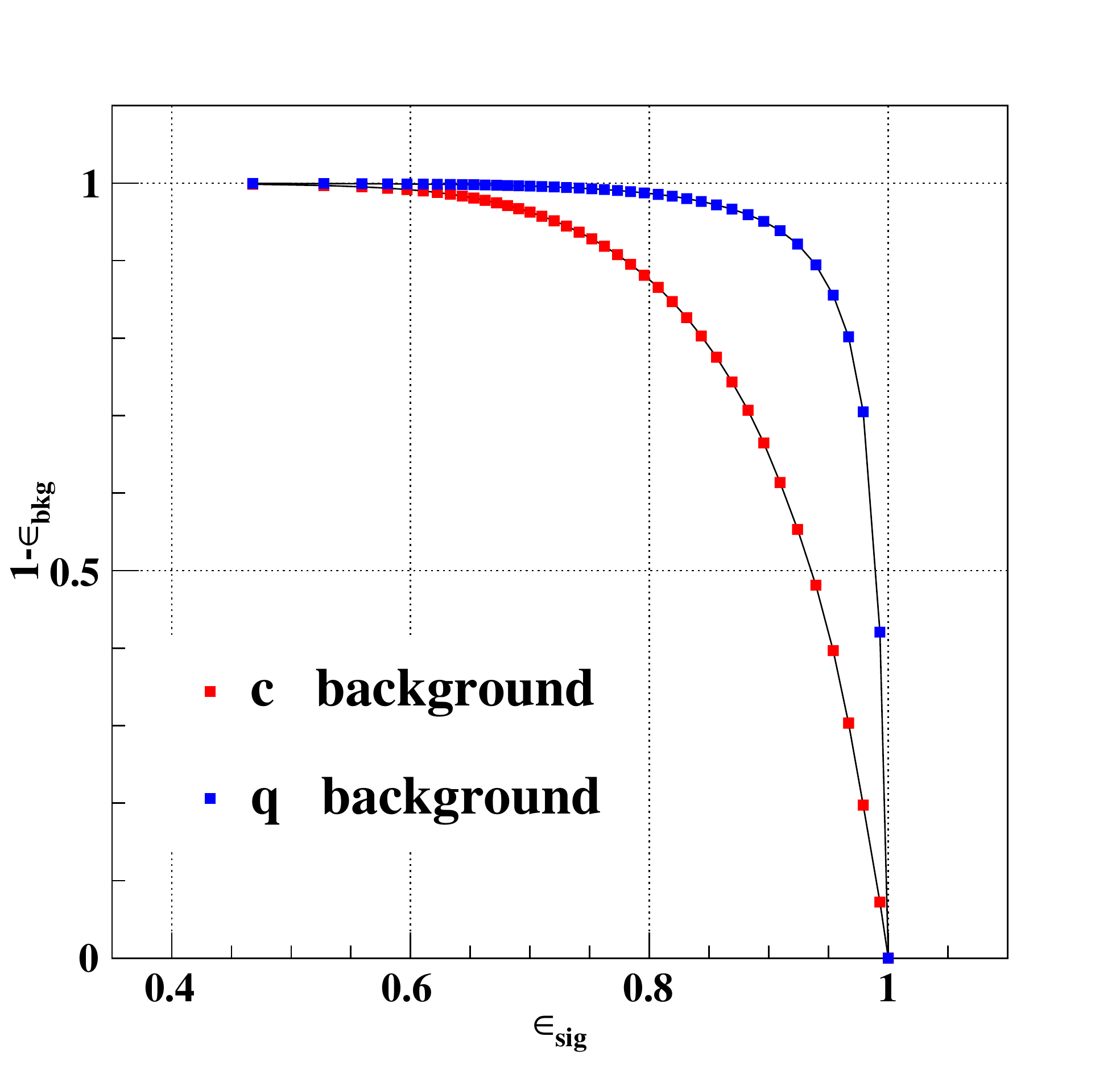}
}
\resizebox{0.4\textwidth}{!}
{
\includegraphics{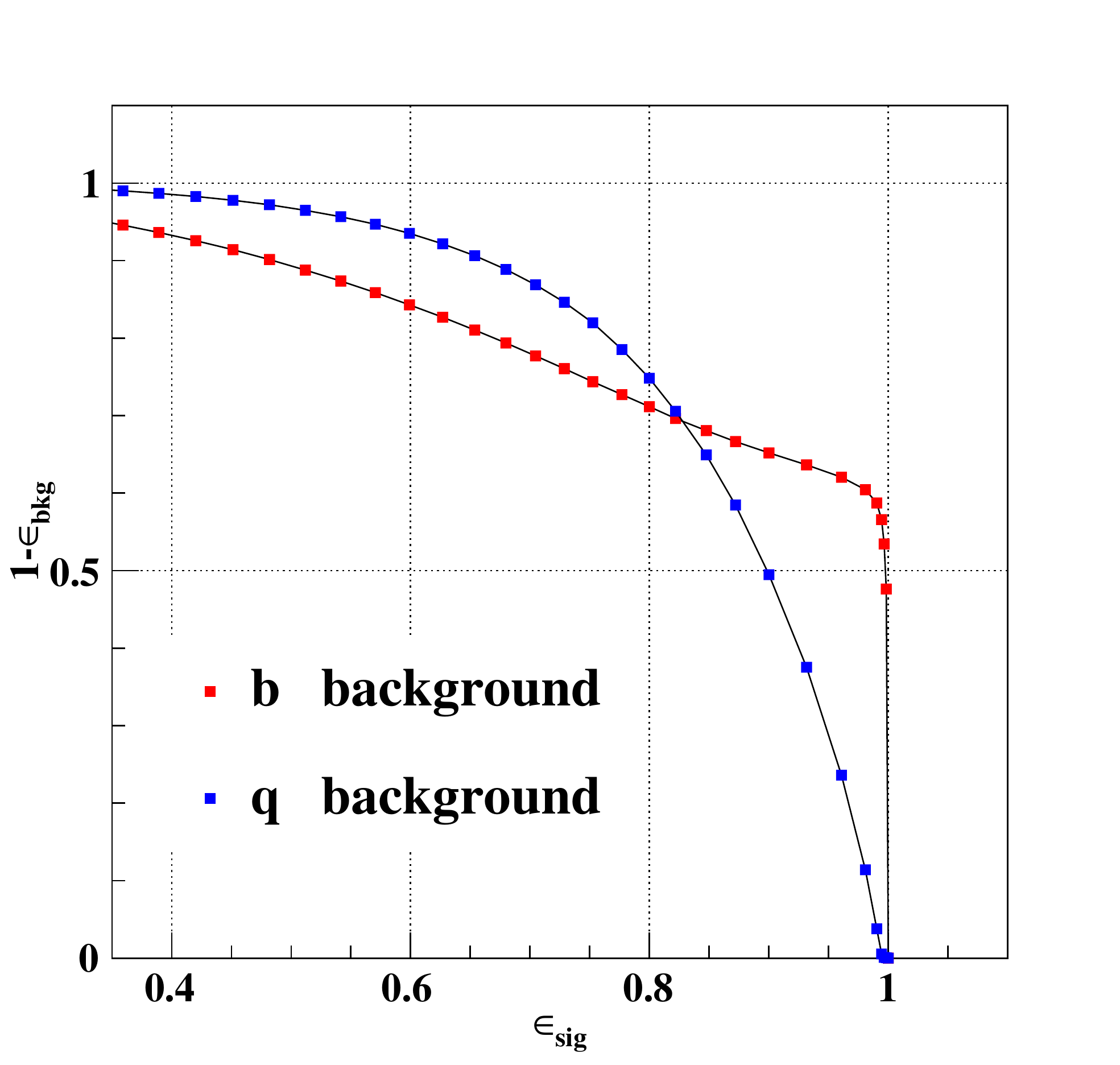}
}
\caption{
  The jet flavor tagging performance.
  }
\label{fig:jetft}
\end{center}
\end{figure}

Technically, the flavor tagging is operated using the LCFIPlus package~\cite{LCFIPlus}, the default flavor tagging algorithm for the linear collider studies. 
At CEPC studies, the LCFIPlus takes the reconstructed final state particles from Arbor, reconstructs the second vertexes and performs the flavor tagging. 
For each jet, LCFIPlus extracts more than 60 distinguish observables and calculates the corresponding b-likeness and c-likeness using the Boost Decision Tree method~\cite{ROOTTMVA}.
Since the b-mesons have longer lifetime compared to the c-mesons, the c-tagging is much more challenging than the b-tagging. 
Thanks to the high precision vertex system, the c-jet could be distinguished from other jets at the ILD detector and the CEPC v\_1 detector. 
Fig.~\ref{fig:jetft} shows the reference ROC curve trained on $Z\to q\bar{q}$ sample at 91.2 GeV center of mass energy. 
The X-axis indicates the b/c-jet efficiency, while the Y-axis represents the surviving rate for the backgrounds.

Applying to the inclusive $Z\to q\bar{q}$ sample, the typical performance of the b-tagging reaches an efficiency/purity of 80\%/90\%, changing the working point to a reduced efficiency of 60\%, the purity could be enhanced close to 100\%. 
While for c-tagging, a typical working point has the efficiency/purity of 60\%/60\%. 

It should be emphasized that, with the current detector geometry design and reconstruction algorithm, the c-tagging is still very difficult. 
As a result, the accuracy of $g(Hc\bar{c})$ measurement is largely limited by the contamination from the $H\to b\bar{b}$ events.

\section{Conclusion}
\label{Summary}

Adequate reconstruction and detector designs are crucial for the success of particle physics experiments. 
Targeting at precise the precise measurements of the Higgs boson properties and the EW observables, the CEPC needs detectors that can reconstruct all the physics objects generated at its Higgs/EW events.
The current CEPC studies use Arbor reconstruction and the PFA oriented detector designs as the baseline. 
This manuscript provides a global description of the physics performance on the physics objects reconstruction and on some benchmark analyses.

Arbor is optimized to fulfill the CEPC physics requirements. 
It reads all the calorimeter hits and tracks and builds reconstructed particles. 
The physics objects are then reconstructed from the reconstructed particle list. 
Inspired by the tree topology of the particle showers, Arbor could efficiently separate nearby particle shower, reconstruct the inner shower structure, and maintain a good energy collection efficiency for individual particles. 
Applying Arbor at the CEPC v\_1 geometry, the following performance has been achieved. 

\begin{itemize}
\item[]1, Lepton identification: $\epsilon_{e \to e} >99.5\%$, $\epsilon_{\mu \to \mu} >99.5\%$, $P_{h \to lepton} <1\%$ for isolated tracks with energy larger than 2~GeV; 

\vspace*{0.3cm}

\item[]2, Charged Kaon identification: efficiency/purity of 91-97\%/94-97\% at inclusive Z pole sample with energy range of 2 - 20 GeV;

\vspace*{0.3cm}

\item[]3, Photon reconstruction: a relative accuracy of 1.7\%/2.3\% is achieved for the Higgs mass reconstruction at $H\to\gamma\gamma$ event using simplified/CEPC v\_1 detector geometry;

\vspace*{0.3cm}

\item[]4, $\tau$: A relative accuracy of 1\% could be achieved for the signal strength measurement of $H\to \tau^+\tau^-$ events;

\vspace*{0.3cm}

\item[]5, Jet energy resolution: A relative accuracy of 3.8\% of Boson mass reconstruction is achieved at a cleaned $H\to gg$ event sample. 
The Higgs boson, the Z boson, and the W boson can be efficiently separated from each other in their hadronic decay modes. 
The jet energy scale is controlled to 1\% level. 
At individual jet, the relative jet energy varies from 3\% to 6\%, depending on the jet transverse momentum.

\vspace*{0.3cm}

\item[]6, Jet Flavor Tagging: at the inclusive $Z\to q\bar{q}$ samples at 91.2 GeV, the b-jets could be identified with an efficiency/purity of 80\%/90\%; while the c-jets could be identified with efficiency/purity of 60\%/60\%.

\vspace*{0.3cm}

\end{itemize}

These key physics objects at the CEPC can be successfully reconstructed. 
The performances at the single particle level, such as the leptons, the kaons, and the photons at simplified geometry, are close to the physics/hardware limits.
The separation and high-efficiency reconstruction of charged particles/photons ensure good $\tau$ lepton reconstruction. 
The jet energy resolution leads to a clear separation between massive bosons at di-jet events.
At individual jets, the uncertainty induced by the final state particle reconstruction is comparable or smaller than these from jet clustering algorithms. 
Meanwhile, using final state particles reconstructed by Arbor, the LCFIPlus algorithm could distinguish b-jet, c-jet, and light-jet from each other. 
In terms of overall performance, the Higgs couplings to its decay final states can be determined to 0.1-1\% accuracy, mostly limited by statistics~\cite{CEPCPreCDR}. 
Therefore, the PFA oriented detector design and Arbor fulfill the CEPC physics requirements on the physics object reconstruction. 

In terms of the reconstruction algorithm development and the detector design, 
huge efforts are needed to bridge the Proof of Principle to the engineering design.
Here we would like to emphasize a few key topics to be explored in the future. 

\begin{itemize}
\item[]1, The systematic control and in-situ monitoring method. 
Systematic control is fundamental to the physics measurements. 
Given the large integrated luminosity at the CEPC, the stability and the systematic control of the CEPC detector system is extremely important and challenging, especially for the Z pole operation. 

\vspace*{0.3cm}

\item[]2, A global design of the DAQ system. 
A global design of the DAQ system, with which the power consumption could be better estimated, is crucial for the further design/optimization work at the detector geometry.

\vspace*{0.3cm}

\item[]3, Detector integration studies.
The detector design needs to ensure that at the integration level, the detector is stable enough to be operated continuously for decades. 
Thermal simulation and mechanic studies are crucial, which have not been covered yet. 
An on-line system that monitors the tension, the temperature, and possibly other condition data like B-field strength, needs to be designed and validated. 

\vspace*{0.3cm}

\item[]4, Development and validation of sub-detector digitization algorithms.
A proper modeling of the detector response is crucial for the systematic control. 
In principle, all the sub-detectors need to have mature test beam references. 
The difference between test beam data and the MC simulation needs to be quantized, properly modeled, and integrated into future simulation tools.

\vspace*{0.3cm}

\item[]5, Advanced reconstruction algorithm and pattern recognition studies.
The current Arbor uses only the hit spatial information in its topological clustering. 
A better usage of the hit time, energy information should significantly enhance its physics performance. 
The pattern recognition plays an essential role in the reconstruction/analysis. 
Meanwhile, the artificial intelligence is in a blooming development. 
The experimental particle physics should also benefit from this trend, making synergies and extend the physics potential accordingly. 

\end{itemize}

\begin{acknowledgement}

We thank Yuanning Gao, Shan Jin, and Zhiqing Zhang for their discussion and input for the CEPC detector study. 
We appreciate the continuous support from in2p3, France, and CAS, China. 
We are deeply grateful to thank Patrick Janot, Luca Malgeri, and other PFA experts at CERN for their physics insight and support during the development of this algorithm. 
We are indebted to our colleagues who worked devotedly on the high granularity calorimeter and future collider studies. 

This work was supported by National Key Program for S\&T Research and Development (Grant No.: 2016YFA0400400), 
the National Natural Science Foundation of China (Grant No.: 11675202), 
the Hundred Talent programs of Chinese Academy of Science (Grant No.: Y3515540U1),
the general research program of Taiwan (Grant No.: MOST-106-2112-M-001-023),
the grant of Ta-You Wu memorial award (Grant No. 105-2112-M-008-022-MY3), 
and the H2020 project AIDA-2020 (Grant No.: 654168).

\end{acknowledgement}

%
%
%
%

\end{document}